\documentclass[journal=ancac3,manuscript=article]{achemso}

\usepackage{chemformula} 
\usepackage[T1]{fontenc} 
\usepackage{hyphenat} 
\usepackage{amsmath,amssymb}
\usepackage{bbding}
\usepackage[notextcomp]{stix}

\author{Jacopo Vialetto}
\affiliation{Laboratory for Soft Materials and Interfaces, Department of Materials, ETH Z{\"u}rich, Vladimir-Prelog-Weg 5, 8093 Z{\"u}rich, Switzerland}
\email{jacopo.vialetto@mat.ethz.ch}

\author{Fabrizio Camerin}
\affiliation{CNR Institute for Complex Systems, Uos Sapienza, P.le A. Moro 2, 00185 Roma, Italy}
\alsoaffiliation{Department of Basic and Applied Sciences for Engineering, Sapienza University of Rome, via A. Scarpa 14, 00161 Roma, Italy}
\email{fabrizio.camerin@gmail.com}

\author{Fabio Grillo}
\affiliation{Laboratory for Soft Materials and Interfaces, Department of Materials, ETH Z{\"u}rich, Vladimir-Prelog-Weg 5, 8093 Z{\"u}rich, Switzerland}

\author{Lorenzo Rovigatti}
\affiliation{Department of Physics, Sapienza University of Rome, P.le A. Moro 2, 00185 Roma, Italy}
\alsoaffiliation{CNR Institute for Complex Systems, Uos Sapienza, P.le A. Moro 2, 00185 Roma, Italy}

\author{Emanuela Zaccarelli}
\affiliation{CNR Institute for Complex Systems, Uos Sapienza, P.le A. Moro 2, 00185 Roma, Italy}
\alsoaffiliation{Department of Physics, Sapienza University of Rome, P.le A. Moro 2, 00185 Roma, Italy}
\email{emanuela.zaccarelli@cnr.it}

\author{Lucio Isa}
\affiliation{Laboratory for Soft Materials and Interfaces, Department of Materials, ETH Z{\"u}rich, Vladimir-Prelog-Weg 5, 8093 Z{\"u}rich, Switzerland}
\email{lucio.isa@mat.ethz.ch}

\title{The effect of internal architecture on the assembly of soft particles at fluid interfaces}


\definecolor{darkspringgreen}{rgb}{0.09, 0.45, 0.27}

\begin{document}



\begin{abstract}
Monolayers of soft colloidal particles confined at fluid interfaces have been attracting increasing interest for fundamental studies and applications alike. However, establishing the relation between their internal architecture, which is controlled during synthesis, and their structural and mechanical properties upon interfacial confinement, which define the monolayer's properties, remains an elusive task. Here, we propose a comprehensive study elucidating this relation for a system of microgels with tunable architecture. We synthesize core-shell microgels, whose soft core can be chemically degraded in a controlled fashion, yielding particles ranging from analogues of standard batch-synthesized to completely hollow microgels after total core removal. We characterize the internal structure of these particles, their swelling properties in bulk and their morphologies upon adsorption at an oil-water interface via a combination of numerical simulations and complementary experiments. In particular, we confirm that hollow microgels are mechanically stable in bulk aqueous conditions and that the progressive removal of the core leads to a significant flattening of the microgels, which become disk-like particles, at the interface. At low compression, the mechanical response of the monolayer is dominated by the presence of loosely crosslinked polymers forming a corona surrounding the particle within the interfacial plane, regardless of the presence of a core. By contrast, at high compression, the absence of a core enables the particles to deform in the direction orthogonal to the interface. These findings shed new light on which structural features of soft particles determine their interfacial behaviour, enabling new design strategies for tailored materials. 
\end{abstract}

\bigskip

The beauty and the strength of using colloidal particles as building blocks for materials lie in our ability to create different structures and elicit different properties by engineering inter-particle interactions.~\cite{Velev2009,Vogel2015a} By tailoring the strength, range and directionality of interactions, a broad range of crystalline and amorphous materials can be made, with the opportunity of reversible and externally-actuated transitions.~\cite{Leunissen2005,Chen2011a,Vogel2015,Shah2015,Montelongo2017,Vialetto2019,Grillo2020} 

The use of soft, deformable particles adds many powerful degrees of freedom to tailor materials.~\cite{Nayak2005} The coupling between particle architecture and deformability in different environments (\textit{e.g.} in bulk suspensions or confined at interfaces, under different temperature or pH conditions, etc.) enables rich behaviour and functionalities unattainable with mechanically rigid "hard" particles.~\cite{Yunker2014,Rey2020}

Microgels, colloidal particles comprising an internally cross-linked network formed by a water-soluble polymer, have emerged as a prominent choice with facile synthesis and versatile applications.~\cite{Pich2010,Plamper2017,Karg2019} The vast majority of microgels are prepared by one-pot precipitation polymerization, which leads to the formation of particles with a radially decreasing cross-linking density profile in bulk aqueous suspensions.~\cite{Bergmann2018,ninarello2019modeling} However, nowadays various synthetic procedures allow realizing microgels with different architectures, such as homogeneously cross-linked~\cite{Acciaro2011,Still2013} or hollow ones~\cite{Zha2002,nayak2005hollow,wypysek2020tailoring,Singh2007,zhang2008preparation,Geisel2015}. The latter particles are made of a polymeric shell surrounding an empty core, and are of particular interest for drug-delivery applications~\cite{bysell2011microgels,kabanov2009nanogels,malmsten2010biomacromolecules}, as well as for their responses to swelling and compression in three dimensions (3D).~\cite{Schulte2018,Nickel2019,Scotti2019a}

Nonetheless, reaching a systematic and comprehensive knowledge, which links synthesis procedures and the resulting particle architectures to the single-particle properties and their collective behavior in dense microgels' suspensions still remains elusive. Establishing this link is particularly challenging for the case of microgel particles confined at fluid, either air-water or oil-water, interfaces. This situation is highly desirable for the creation of advanced materials, \textit{e.g.} responsive foams and emulsions, but presents additional issues.~\cite{Karg2019,Rey2020} Even if the 3D architecture of the microgels is controlled during the bulk synthesis processes, as the particles adsorb to a fluid interface, surface tension and the different solubility of the polymer in the two phases cause a reconfiguration of the polymer network.~\cite{Geisel2012,Geisel2015,camerin2019microgels,Scotti2019} Typically, standard microgels tend to flatten out at the interface, with the more loosely crosslinked polymer chains at the particle periphery creating a thin corona, which surrounds the particle within the interfacial plane. Currently, it is still unclear how the 3D, bulk shape of the microgel translates into its new shape at the interface and how the mechanical properties of the particle in bulk are transferred to the interface. Numerical and experimental studies have shown that steric interactions of microgels in bulk and at interfaces can be described using different kinds of Hertzian elastic models~\cite{bergman2018new,Grillo2020,camerin2020microgels}. Moreover, the characteristic swelling-deswelling behavior of the microgels in bulk translates into different mechanical behavior at interfaces as a function of compression, temperature and pH, which is yet to be rationalized from single particle properties and interactions.~\cite{Geisel2014,Bochenek2019,Harrer2019} 

In this manuscript, we address this question by realizing a system of core-shell microgels, where the inner microgel core can be chemically degraded in a controlled fashion,~\cite{nayak2005hollow} to tune the particle architecture from the one of a standard, batch-synthesized microgel to a hollow one. The progressive removal of the core can be followed both experimentally and numerically and it strongly affects both the swelling-deswelling transition in bulk and the morphology at the interface. We present our outcomes by first examining the characteristics of individual particles in bulk and at oil-water interfaces as a function of core removal, and then focusing on the collective response of hollow microgels under compression at the interface. Here, we disclose a rich mechanical and structural behavior, which sheds new light on future design routes to obtain tailored two-dimensional assemblies of soft colloids.

\section{Results and discussion}

\subsection{Characterization of the hollow microgels in bulk}

\subsubsection{Core degradation}

\begin{figure}[t!]
\centering
\includegraphics[scale=0.89]{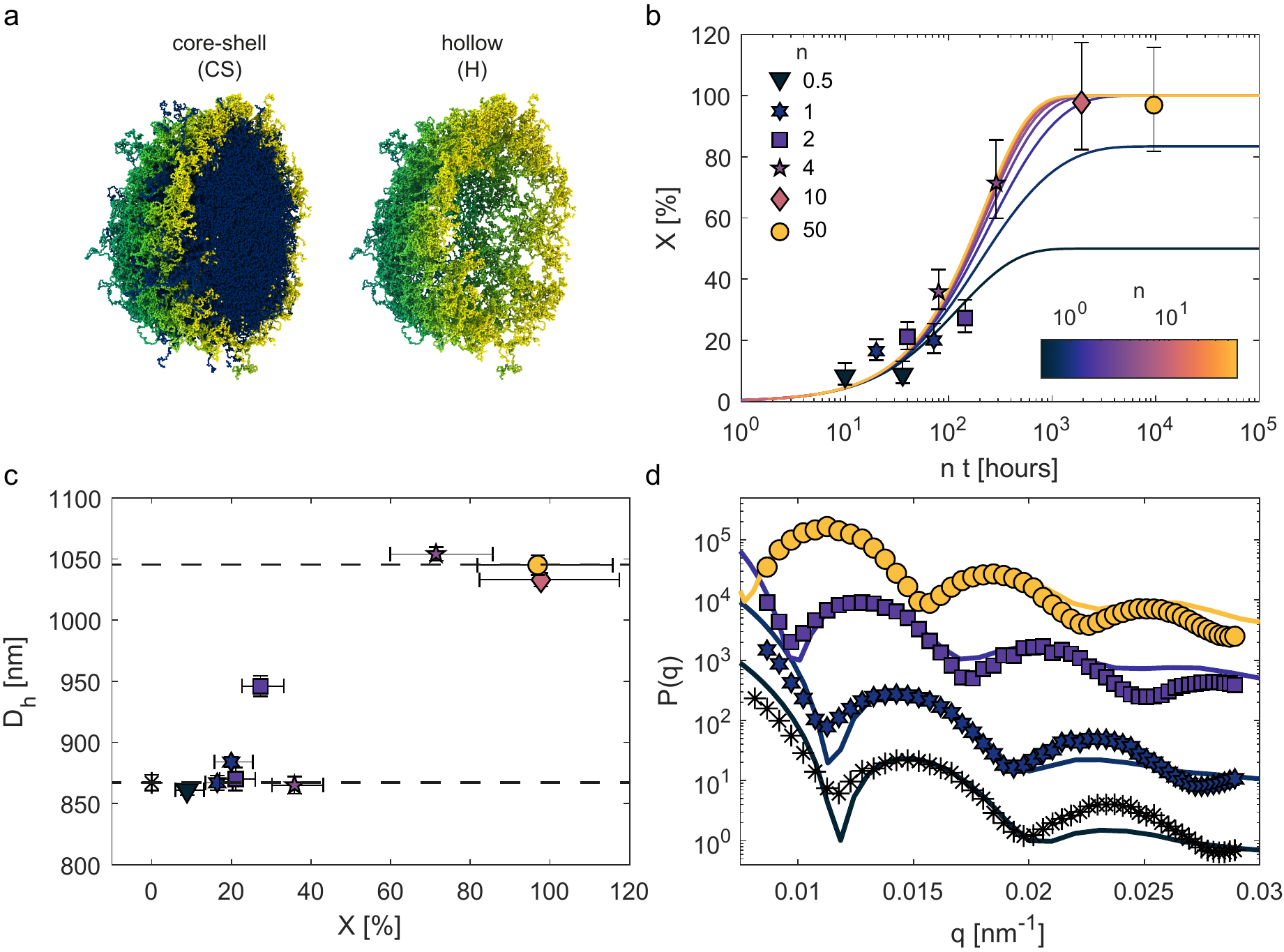}
\caption{\small \textbf{Core-shell and hollow microgels in bulk aqueous suspensions: core degradation process.} (a) Simulation snapshots of the cross-section of a core-shell microgel before and after the removal of the inner core. The core with the DHEA crosslinker is colored in blue and the outer shell with the BIS crosslinker is colored in green/yellow. (b) Extent of core removal $X$ as a function of $n \cdot t$, where $n$ is the NaIO$_4$ to DHEA molar ratio and $t$ the reaction time. The symbols indicate the values of $X$ estimated from the experimental form factors (see Methods), whereas the lines show Equation \ref{eq:core_removal} fitted to the experimental $X$ for different values of $n$. (c) Hydrodynamic diameter $D_h$ of core-shell microgels as function of the extent of core removal $X$. The horizontal dashed lines emphasize the sudden transition to a larger dimension upon removal of a critical core fraction. (d) Experimental (symbols) and numerical (lines) form factors as a function of the wavenumber $q$. The experimental form factors are for $n \cdot t=0, 20, 144, 9600$, while the numerical ones are calculated for different number densities of the core microgel: $\rho \approx 0.08, 0.06, 0.01, 0 ~ \sigma^{-3}$ (from the bottom to the top), with $\rho=0$ corresponding to the hollow microgel ($X=100\%$). The form factors are arbitrarily shifted along the $y$-axis for visual clarity. The symbols in (c) and (d) correspond to the same values of $n$ shown in (b), with $\ast$ indicating the core-shell microgel prior to core removal ($n=0$).}
\label{fig:Figure1}
\end{figure}

The hollow microgels used in this study are synthesized by means of a two-step polymerization procedure, followed by a degradation process.~\cite{nayak2005hollow} We first obtain a standard 
microgel by free-radical precipitation polymerization using DHEA as crosslinker, and afterwards we grow a second polymeric shell on top using BIS as crosslinker, forming a \textit{core-shell} particle (see Methods for a detailed description of the synthesis).
By exposing the core-shell microgels to NaIO$_4$, the DHEA crosslinkers in the core can be degraded, leading to the progressive formation of \textit{hollow microgels}, as represented by the simulation snapshots of Fig.~\ref{fig:Figure1}(a). 

We study the kinetics of the degradation of the core as a function of the time $t$ and the initial NaIO$_4$ to DHEA molar ratio $n=\frac{\left[{\rm NaIO_4}\right]_0}{\left[{\rm DHEA}\right]_0}$ via \textit{ex-situ} SLS and DLS measurements.
In particular, we vary $n$ in the range 0.5 - 50 and $t$ in the range 20 - 192 hours and define the degree of core removal as $X(n,t)=\frac{M_c^0-M_c(n,t)}{M_c^0}$.
Here, $M_c^0$ and $M_c(n,t)$ are the mass of the core before and after the reaction with NaIO$_4$, respectively. These are estimated from the microgels' radial density profiles, which we extract by fitting the form factors for different values of $n$ and $t$, as described in Methods and reported in Figs. S1 and S2.
The symbols in Fig.~\ref{fig:Figure1}(b) represent the experimental $X(n,t)$ for different values of $n$. When plotted against the product $n\cdot t$, $X$ exhibits a sigmoidal trend, with larger $n\cdot t$ leading to increasingly higher $X$ up to $n\cdot t\simeq 10^3$ hours after which $X$ reaches a plateau at $\simeq 100$\%.

To rationalize such kinetics, we model the degradation of the core as a bimolecular and stoichiometric reaction between the vicinal diol group in the DHEA polymer network and NaIO$_4$~\cite{nayak2005hollow}. The reaction $\rm DHEA_{(s)}+{\rm NaIO_4}_{(l)}\longrightarrow C_{(l)}$ leads to a soluble product $C_{(l)}$ that is washed out during the post-reaction cleaning step. By assuming the reaction to follow a first-order kinetics with respect to the concentration of both DHEA and NaIO$_4$, perfect mixing within the reaction vessel, and a one-to-one linear correlation between the degree of core removal and the cleavage of the vicinal diol in the DHEA molecules, $X$ is prescribed by the following differential equation:

\begin{equation}
    \frac{dX}{dt}=k \left[{\rm DHEA}\right]_0\left(1-X\right)\left(n-X\right)
\end{equation}
where $k$ is the reaction rate constant. This equation admits two analytical solutions depending on whether $n=1$ or $n \neq 1$:
\begin{equation}
X(n,t)=\begin{cases}
               \frac{n\left(e^{\left(n-1\right)\left[{\rm DHEA}\right]_0kt}-1\right)}{n~e^{\left(n-1\right)\left[{\rm DHEA}\right]_0kt}-1},n \neq 1\\
               \frac{\left[{\rm DHEA}\right]_0kt}{\left[{\rm DHEA}\right]_0kt+1},~n = 1.
\end{cases}
\label{eq:core_removal}
\end{equation}

It is important to remark that, within this framework, for a given initial DHEA concentration $\left[{\rm DHEA}\right]_0$, varying $n$ is not equivalent to varying $t$ as $X(n,t)$ is not a function of the product $n\cdot t$. This emphasises the importance of studying the effects of varying both $n$ and $t$ separately. Nonetheless, for $n>>1$, $X(n,t)$ can be approximated to
a sole function of the product $n\cdot t$ as $X(n,t)\sim (e^{nkt}-1)/e^{nkt}$. 
The fitting of the experimental $X(n,t)$ with Eq.~\ref{eq:core_removal} is plotted in Fig.~\ref{fig:Figure1}(b) against $n\cdot t$, giving an excellent description of the experimental data ($R^2=0.96$) with a value of $k=1.33\cdot 10^{-2} \pm 4 \cdot 10^{-3}$ m$^3$ mol$^{-1}$ hour$^{-1}$. 

The degree of core removal correlates with the microgels' hydrodynamic diameter $D_h$, as measured by DLS. The standard, DHEA-crosslinked (core) microgels prior to the growth of the BIS shell have $D_h =640$ nm at 22°C (blue triangle in Fig. S3). After the second polymerization step, the $D_h$ of the core-shell particle reaches 870 nm (22°C). We then calculate the thickness of the BIS shell as $(D_h(CS) - D_h(C))/2$, obtaining a value of 115 nm. After reaction with NaIO$_4$, $D_h$ remains roughly constant up to $X \simeq 40 \%$ after which it rapidly increases, reaching a plateau at $\simeq 1050$ nm for $X \simeq 70 \%$, as schematically represented by the two dashed horizontal lines in Fig.~\ref{fig:Figure1}(c).
This is in agreement with the plateau observed for the degree of core removal in Fig.~\ref{fig:Figure1}(b) and indicates that the dissolution of the core allows for an increased swelling of the pNIPAM network in the shell. 

To further support these findings, we perform a study in which we verify the compatibility of the form factors extracted experimentally at different degradation stages with those calculated from a numerical model for hollow microgels. The model that best reproduces the experimental data at the highest $X$ is determined by adjusting the size of the inner cavity and the monomer density of the polymeric shell in the simulations. A faithful comparison is achieved, as shown in Fig.~\ref{fig:Figure1}(d), for a hole radius that comprises $75\%$ of the microgel diameter (compared to $74 \pm 2\%$ from the DLS measurements) and an outer shell having an average monomer density of about half that of the standard microgel used for the core particle (see Methods and Supporting Information, Figs. S7-S10).

Since it is not possible to reproduce the chemical degradation procedure numerically, we mimic this process by taking the hollow model just described, inserting a standard microgel in the central cavity (representing the inner DHEA-crosslinked microgel) and by subsequently removing a certain number of monomers from the inner core, leading to a decrease of its density (see Methods and Fig. S11). As shown in Fig.~\ref{fig:Figure1}(d), the analysis of the form factors confirms the trend we observe in experiments as a function of $X$, with the first peak of the form factor shifting to smaller wavenumbers and the simultaneous appearance of the third peak in the range of examined wavenumbers.

\subsubsection{Temperature response}

Numerical simulations offer us further insights on the internal structure of the microgels upon core degradation  by studying their temperature response. In order to substantiate the numerical results with experiments, we perform DLS and SLS measurements at different temperatures below and above the pNIPAM's volume phase transition temperature (VPTT) of 35 °C. Above the VPTT (40 °C), when pNIPAM is in a collapsed state, $D_h$ decreases from 520 nm (for X = 0\%) to 490 nm (for X $\simeq$ 100\%, Fig. S3). 
This indicates that, upon degrading the core particle, the outer shell does not buckle into the particle interior, despite being very thin ($\simeq$ 15 nm). Nonetheless, the swelling ratio $D_h$(22°)/$D_h$(40°) is higher for a hollow microgel (2.1 $\pm$ 0.05) than for the initial core-shell particle (1.66 $\pm$ 0.04) because $D_h$(22°) increases with $n t$ and thus $X$.

\begin{figure}[t!]
\centering
\includegraphics[scale=0.94]{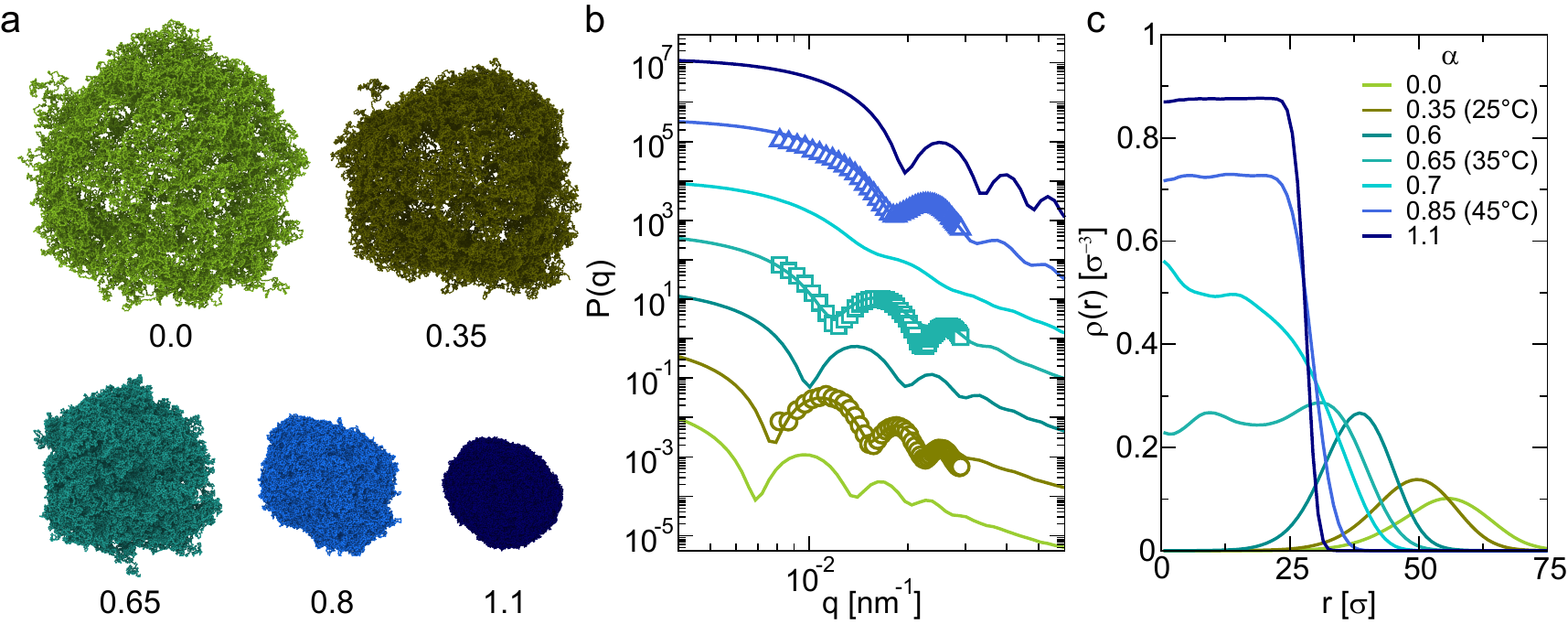}
\caption{\small \textbf{Responsiveness of the hollow microgels across the VPT.} (a) Simulation snapshots of hollow microgels as a function of the effective temperature $\alpha$; (b) Experimental (symbols) and numerical (lines) form factors for different effective temperatures. The values of $\alpha=0.35, 0.65$ and $0.85$ correspond to $ T=25, 35$ and 45°C, respectively. The comparison is performed by matching the positions of the first peak of $P(q)$, yielding an estimate of the bead size used in the simulations $\sigma=8.25$ nm. The form factors are arbitrarily rescaled on the $y$-axis for visual clarity; (c) Density profiles as a function of the distance from the center of mass of the hollow microgel in units of $\sigma$ for different values of the effective temperature $\alpha$.}
\label{fig:hollowvpt}
\end{figure}

The temperature response of the hollow microgels (i.e. with $X\simeq 100\%$) emerging from the SLS measurements is fully captured by the simulations, where the entire range of effective temperatures can be accessed by varying the parameter $\alpha$ of the solvophobic potential (see Methods). Fig.~\ref{fig:hollowvpt}(a) displays representative simulation snapshots, showing the transition from a fully swollen microgel to a collapsed one at $\alpha=1.1$. It is evident by the snapshots that, due to their intrinsic structure and the reduced density of the shell, these hollow microgels appear more anisotropic and fluffy than the corresponding standard microgels.~\cite{ninarello2019modeling}

The alignment with real units is obtained by matching the first peak of the numerical form factor onto the experimental one at T=$25$ °C and by adopting the same rescaling factor for the other temperatures.~\cite{ninarello2019modeling} The comparison, reported in Fig.~\ref{fig:hollowvpt}(b), shows an almost perfect agreement between the form factors of model and laboratory microgels which is thus realized for $\alpha=0.35, 0.65$ and $0.85$ corresponding to T=$25,35$ and $45$ °C, respectively. In particular, we note how the curves progressively shift to higher wavevectors, consistently with the decrease in particle size measured with DLS (Fig. S4). Most importantly, we find a crossover at $\alpha=0.7$, roughly corresponding to $36-37$°C, where the form factor has a smoother shape before exhibiting sharp peaks at higher $\alpha$ values. 

Having established the correspondence between experimental and numerical form factors, we examine the calculated radial density profiles to confirm the creation of hollow microgels below the VPTT. Fig.~\ref{fig:hollowvpt}(c) indeed shows a polymer density profile characterized by a Gaussian-like shell, with a cavity occupying most of the extension of the microgel.  However, from around $35$°C, the cavity in the center of the microgel starts to be filled by the polymer chains. Above the VPTT, the density profiles resemble those of a standard microgel. 

\subsection{Hollow microgels at fluid interfaces}

\subsubsection{Individual microgels}

To ultimately rationalize how a given internal polymeric structure of the microgels affects their 2D assembly at fluid interfaces, we begin by analyzing the conformation that individual particles assume after adsorption.

An indirect measurement of the 3D shape of the microgels at the oil-water interface is obtained by visualizing the colloids after deposition from the liquid interface onto a solid substrate. AFM imaging allows for a precise quantification of the height profiles of dried microgels. Following previous works~\cite{camerin2019microgels,rey2016isostructural,geisel2014highly}, we assume that the profiles measured in this condition correlate with the conformation the microgels had at the interface prior to deposition. 

\begin{figure}[t!]
\centering
\includegraphics[scale=0.95]{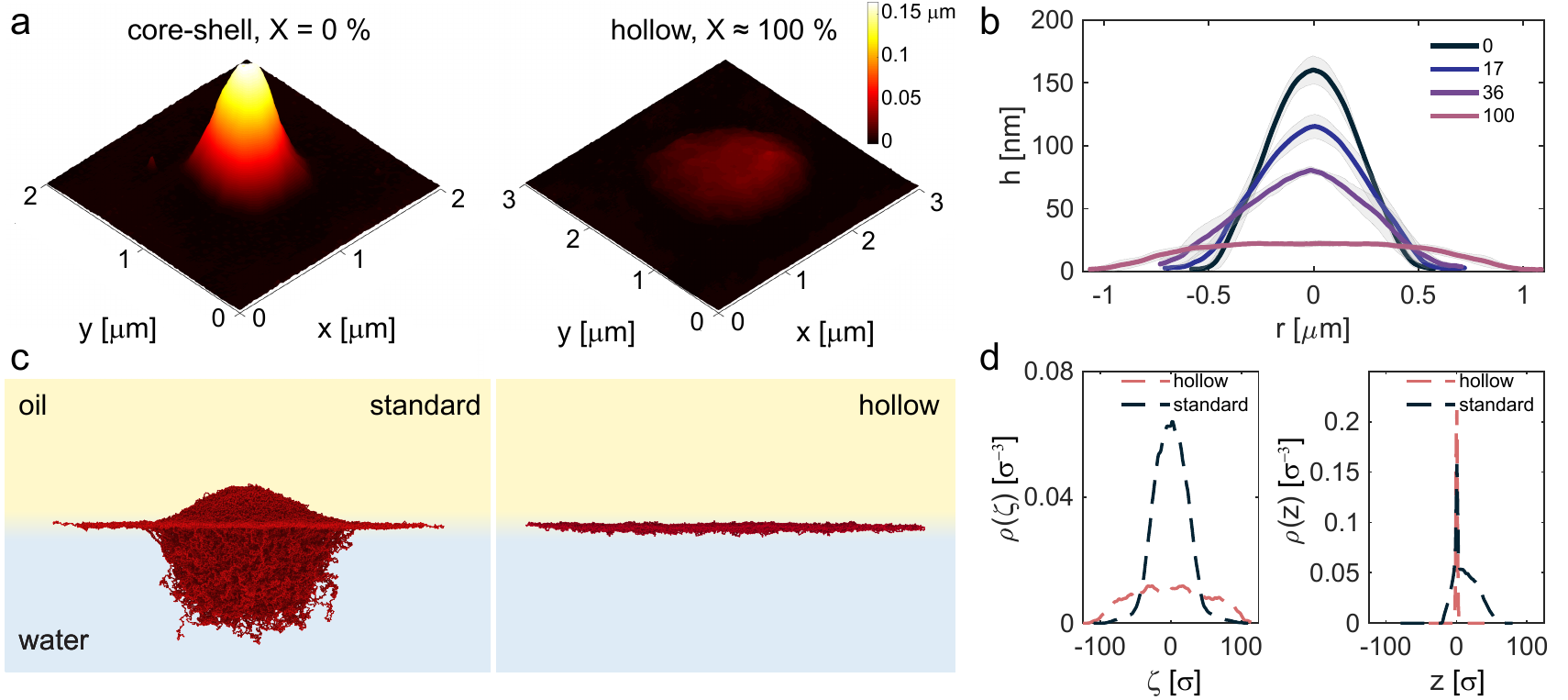}
\caption{\small \textbf{Core-shell and hollow microgels at the liquid interface.} (a) 3D profiles of dried core-shell (left) and hollow (right) microgels as obtained by AFM imaging after deposition onto a silicon wafer from the water/hexane interface. (b) Experimental height profiles for different $X$ [$\%$]. The shaded regions correspond to the standard deviations of the height profiles calculated on around 10 particles. (c) Simulations snapshots of a standard (left) and a hollow (right) microgel adsorbed at a oil-water interface.  (d) Numerical density profiles on the plane of the interface $\rho(\zeta)$ (left)  
and across the interface $\rho(z)$ (right) for a hollow microgel and for a standard core-corona microgel; for $\rho(z)$, negative $z$ corresponds to the oil phase.
}
\label{fig:singlemgelinterf}
\end{figure}

Fig.~\ref{fig:singlemgelinterf}(a) reports typical 3D profiles of isolated microgels deposited on the solid substrate before and after the core degradation process. The reconstructed height profiles at various degree of core degradation are reported in Fig.~\ref{fig:singlemgelinterf}(b). Untreated core-shell microgels (\textit{i.e.}, $X=0\%$) display the characteristically common "core-corona" morphology comprising a denser and thicker core and a more spread-out polymer layer (or corona) of decreasing thickness.\cite{Geisel2012} Upon increasing $X$, the height profiles reveal that the microgel flattens and its extension on the interfacial plane increases. In other words, upon removal of the internal polymer network, the microgels become more deformable and the outer shell, less and less constrained by the reduction of cross-links in the core, can further stretch and expand at the interface. Ultimately, a totally different profile is obtained for the hollow particle ($X \simeq 100 \%$), which takes up a flat disk-like shape at the interface, with a maximum thickness of around 20 nm after deposition. In all cases, an external corona composed by the outermost polymer chains, which spread out on the interfacial plane, is found around the microgels, as it is clearly shown by AFM phase images (see discussion below).

The change in morphology observed during core degradation is different from the effect of reducing the percentage of crosslinkers in standard batch-synthesized microgels. For the latter particles, decreasing the internal crosslinking density also results in an increased spreading at the interface together with a concomitant decrease of the maximum height.\cite{Picard2017,Rey2017} However, in that case, the height profile retains a Gaussian-like shape, which remains overall qualitatively similar for different cross-linking degrees.~\cite{camerin2019microgels}

A closer look at the height profiles for our microgels reveals that their maximum height decreases up to half the initial height (and the projected area doubles) already for only $\approx 36\%$ of core removal. We hypothesize that, at this stage, a limited number of crosslinks in the internal polymer network are cleaved. While this does not alter significantly the bulk internal density of the polymer composing the particle, it significantly changes its internal elasticity. The particle becomes less stiff and can consequently spread more at the interface under the action of the interfacial tension.

In simulations, we compare the particle morphology of a standard and of a hollow microgel. Simulation snapshots are reported in Fig.~\ref{fig:singlemgelinterf}(c) and immediately illustrate the difference between the two architectures. The microgel density profiles projected onto the plane of the interface $\rho(\zeta)$ (see Methods), reported in Fig.~\ref{fig:singlemgelinterf}(d),
closely mirror the experimentally measured ones after deposition (Fig.~\ref{fig:singlemgelinterf}(b)). In particular, these confirm a similar distribution of the polymer network between core-shell particles and standard microgels, while, for the hollow ones, the numerically extracted shape compares well to that obtained experimentally. In the same figure we also report the profiles across the interface $\rho(z)$. 
Here, as opposed to the case of the standard microgel, which shows different protrusions in oil and water, the hollow microgel is essentially confined within the interface where it forms a uniform thin layer, 
barely protruding into either fluid.

Finally, we characterize the extent of the in-plane deformation at the oil-water interface relative to bulk size by the ratio $D_{i}^0/D_{h}$, where $D_{i}^0$ is the particle diameter at the interface as measured by AFM or extracted from the simulations. The core-shell microgels have a stretching ratio of $1.20 \pm 0.04$, while the hollow microgels reach a value of $1.8 \pm 0.1$, compatible with the values extracted from the numerical simulations (1.4 and 1.9, respectively).

\subsubsection{2D assembly and response upon compression}

After examining their single-particle conformation at the oil-water interface, we study 2D assemblies of the hollow microgels to link their architecture with their 2D phase behaviour and response upon interfacial compression. To this purpose, we spread microgels at a water/hexane interface in a custom-built Langmuir trough and transfer monolayers under continuous compression as detailed in the Methods section.

\begin{figure}[t!]
\centering
\includegraphics[scale=0.95]{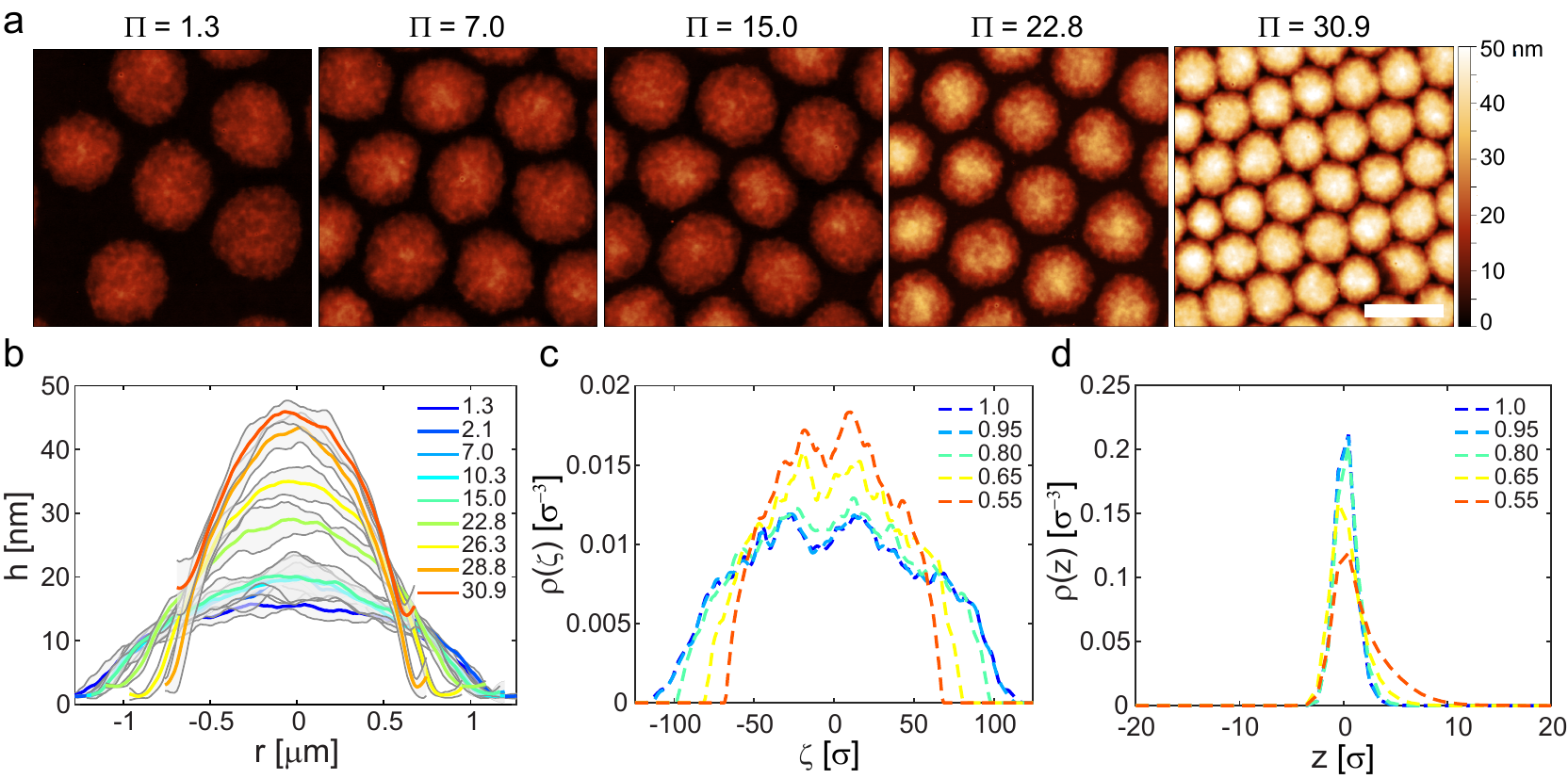}
\caption{\small \textbf{Compression of hollow microgels at the liquid interface.} (a) AFM height images showing the monolayer microstructure at increasing surface pressure $\Pi$. Scale bar: 2 $\mu$m. (b) Microgel height profiles extracted from images as in (a) at increasing $\Pi$. The shaded regions correspond to the standard deviations of the height profiles calculated on around 6 to 10 particles. (c-d) Numerical density profiles within the plane of the interface $\rho(\zeta)$ (c) and across the interface $\rho(z)$ (d), for different degrees of radial compression of the hollow microgels defined as $D_{i}/D^0_{i}$, where $D^0_{i}$ is the interfacial size of hollow particle without compression; for $\rho(z)$ negative $z$ corresponds to the oil phase.}
\label{fig:collectivemgelinterf}
\end{figure}

\begin{figure}[t!]
\includegraphics[scale=0.95]{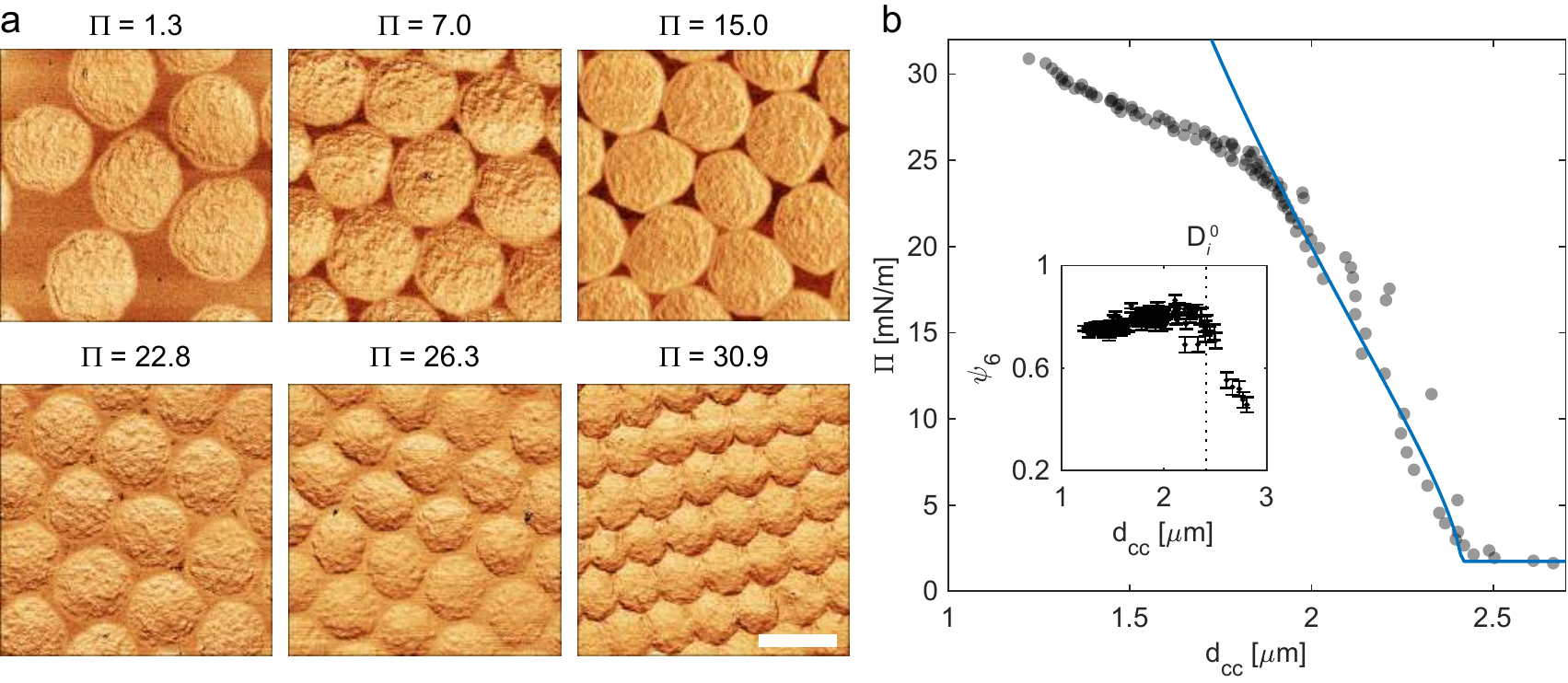}
\caption{\small \textbf{Collective behaviour of hollow microgels at the oil-water interface.} (a) AFM phase images showing the microgels' conformation at different values of the surface pressure $\Pi$. Scale bar: 2 $\mu$m. (b) Compression isotherms reporting the surface pressure $\Pi$ \textit{versus} the center-to-center distance $d_{cc}$ in the particle monolayer. The first part of the compression curve is fitted to Eq.~\ref{eq:GHP}. Inset: hexatic order parameter $\Psi_6$ as a function of $d_{cc}$. The black dotted line indicates the diameter of a single particle at the liquid interface prior to compression ($D^0_{i}$).
}
\label{fig:compressionmgelinterf}
\end{figure}

Fig.~\ref{fig:collectivemgelinterf} reports the microgels' conformation within a monolayer upon increasing compression of the liquid interface, as revealed by AFM height images. At low pressures (up to $\Pi \approx 15 mN \cdot m^{-1}$), the microgels deform in a limited manner (see height profiles in Fig.~\ref{fig:collectivemgelinterf}(b)): both the cross-sectional area within the interface plane and the height of each microgel remain approximately constant. In this compression range, microgels organize into a hexagonal assembly and show no appreciable deformation. As we will discuss later, we hypothesize that increasing compression in this regime mostly affects the outermost polymer chains, which occupy all the available space in between the particles. These observations are also captured by the numerical simulations results reported in Fig.~\ref{fig:collectivemgelinterf}(c-d). Here, the radial compression is progressively carried out on single hollow microgels (see Methods). The numerical density profiles, both on the plane of the interface (Fig.~\ref{fig:collectivemgelinterf}(c)) and across the interface (Fig.~\ref{fig:collectivemgelinterf}(d)), are essentially not affected if $D_{i}/D_i^0 \lesssim 5\%$ (corresponding to $\Pi \lesssim 15 mN \cdot m^{-1}$ in the experiments), resembling in all respects the profile of an uncompressed particle.
However, at higher $\Pi$, the rearrangement of the polymeric network involves the whole monolayer. For $\Pi > 20mN \cdot m^{-1}$, the hollow microgels significantly deform to accommodate the increased pressure: the cross-sectional area decreases, while the height of the deposited microgels increases (Fig.~\ref{fig:collectivemgelinterf}(b)). At this level of compression, part of the polymer chains desorb from the interface and the particle expands in the third dimension, perpendicularly to the interface. 
This extension in the third dimension is clearly visible in the $\rho(z)$ profiles of the simulated microgels, reported in Fig.~\ref{fig:collectivemgelinterf}(d).
Furthermore, as more clearly evidenced from phase images shown in  Fig.~\ref{fig:compressionmgelinterf}(a), the particles now deform into hexagons to fully occupy the available space at the interface.

Information on the macroscopic response of the microgel monolayer comes from the compression curve reported in Fig.~\ref{fig:compressionmgelinterf}(b), which displays the surface pressure as a function of the microgel-microgel interparticle distance $d_{cc}$. It can be noted that the curve presents three different slopes, depending on the range of $d_{cc}$ analyzed.
The first regime at very low and almost constant surface pressure ($d_{cc} > 2.4 \mu m$) corresponds to a dilute monolayer, where the distance between particles is higher than the size of a single particle at the interface, and the result is a disordered arrangement (Fig.~\ref{fig:compressionmgelinterf}(a), $\Pi=1.3 mN \cdot m^{-1}$). 
Compressing such a dilute monolayer causes a progressive increase of the effective concentration of microgels at the interface until all particles in the assembly are in contact through their outer polymeric chains. This happens when the distance between microgels is approximately equal to their lateral size at the interface, marked in Fig.~\ref{fig:compressionmgelinterf}(b) as $D_{i}^0$. Correspondingly, the particles assemble into a hexagonal packing, as evidenced by the gradual increase in the hexagonal order parameter $\Psi_6$, which reaches a high-valued plateau  around $D_{i}^0$ (inset in Fig.~\ref{fig:compressionmgelinterf}(b), corresponding AFM image in Fig.~\ref{fig:collectivemgelinterf}(a) and Fig. S5). 

Upon increasing compression, the monolayer enters a second regime where the surface pressure rises while the particle morphology does not exhibit significant changes, as previously described. Correspondingly, the monolayer retains a high degree of hexagonal order and $\Psi_6$  substantially remains constant.

The effective mechanical behavior of the monolayer for small particle deformations is captured by assuming that the microgels interact via a generalized Hertzian potential.\cite{Grillo2020}. From this description, we extract two important facts. First, the fit nicely describes the data only for values of $d_{cc}$ for which the height profiles of the microgels do not exhibit significant deformation in the third dimension. Second, the fit gives a power-law exponent $a = 1.7 \pm 0.1$, which is markedly similar to the one obtained for common core-corona microgels at the water/hexane interface ($a = 1.8 \pm 0.2$).\cite{Grillo2020} These two facts indicate that the macroscopic response of the monolayer in this compression range is similar, irrespective of the internal microgel structure. As a conclusion, we hypothesize that only the outer pNIPAM chains determine the mechanical behavior upon compression in this surface pressure range.

Finally, we observe a third regime in the compression curve. For $d_{cc} < 2 \mu m$ and $\Pi > 20 mN \cdot m^{-1}$, the out-of-plane deformation of the microgels translates into a variation of the slope of the compression isotherm. 
At first sight, the slope change may resonate with the similar trend that has been reported for batch-synthesized microgels.~\cite{rey2016isostructural,Scheidegger2017} However, in the latter case, for microgels of similar size, this is typically ascribed to an isostructural solid-solid phase transition which occurs when the outer microgel corona collapses enabling core-core contacts.~\cite{rey2016isostructural} In the case of hollow microgels instead, the absence of polymer in the core allows the particles to shrink in the interfacial plane without affecting their hexagonal packing. 
On the macroscopic level, this results in a 2D assembly that maintains the same hexagonal order throughout the compression isotherm, but with a continuously decreasing particle size and, correspondingly, lattice constant (Fig. S6).

\section{Conclusions}

In this work, we coupled experiments and numerical simulations to provide a detailed description of the internal polymer distribution of core-shell microgels undergoing varying degrees of core degradation. In particular, by endowing the soft cores with chemically cleavable crosslinks, we obtained particles with controlled internal structures, ranging from the as-synthesized core-shell case to hollow microgels with a thin shell ($\sim$ 115 nm) and a large cavity ($\sim$ 640 nm), after full degradation of the core. Such architectures, encompassing small ratios of polymer shell thickness versus size of the cavity, are not easily accessible by common synthetic strategies, where hollow particles are made by dissolving a hard inorganic core.~\cite{Nickel2019,Geisel2015}
We found that the internal structure of the particles, which we characterize in the bulk aqueous phase, is tightly linked to both the global mechanical response of microgel monolayers at a fluid interface and their microstructure upon isothermal compression. 
In particular, we rationalized the properties of the two-dimensional assemblies by characterizing the morphology of individual microgels at the fluid interface. Here, we evidenced that the monolayer response at low compression is similar for all microgel internal structures, and appears to be solely governed by the rearrangements of the outer polymer chains constituting the microgel corona at the interface. Conversely, a greater degree of compression led to a different phase behavior of the hollow microgels compared to that of common batch-synthesized ones, with a continuous compaction of a hexagonally packed monolayer instead of the commonly occurring isostructural transition found for standard microgels~\cite{rey2016isostructural}. Similar continuous transitions have been already reported in the case of nm-sized microgels,~\cite{Scheidegger2017} or for microgels with a low cross-linking density~\cite{Rey2017,Picard2017}. However, in these cases, the hexatic order usually drops upon increasing compression, leading to a disordered monolayer at high pressure. The absence of a core instead enables the hollow microgels to expand out of the interface plane, allowing the assembly to maintain a high degree of hexatic order throughout the compression isotherm. Incidentally, the behavior observed for bulk suspensions of hollow microgels is radically different from that described here at the interface, indicating a central role of the surface tension in determining their collective behavior. Indeed, a recent study has shown the absence of crystal formation in three dimensions in favor of deswelling, interpenetration and faceting depending on the concentration regime under investigation.~\cite{scotti2021absence}

Our findings shed new light on the interplay between the internal structure and the interfacial behavior of soft colloids. In particular, we remark that the fact that the structural and mechanical behaviour at the fluid interface under low compressions, \textit{i.e.} corresponding to small overlaps between the particles, appears to be independent of the presence of an inner cavity presents us with new challenges and opportunities during synthesis. Most synthetic efforts are targeted to control the radial cross-linking density profiles of the microgels. However, the interfacial response is initially governed by the properties of a corona of loosely crosslinked or even un-crosslinked chains that surround the microgel within the interface plane. Based on these considerations, we expect that a bigger impact in the tailoring of the monolayer response can be achieved by engineering the properties of these coronas, rather than of the overall density profiles. Possible strategies encompass the controlled growth of shells of linear polymers of different length after completing the crosslinking reaction. At the same time, a closer inspection to the fate of the chains composing the corona during compression would certainly greatly add to our understanding. Nonetheless, the importance of the internal structure emerges at higher compression, delineating a large, and only partially explored, parameter space for the controlled designed of tailored two-dimensional assemblies of soft particles. 

\section{Materials, models and methods}
\small
\subsection{Synthesis of core-shell and hollow microgels}

The hollow microgels used in this study were synthesized by a two-step free-radical precipitation polymerization following, with some modifications, a procedure previously reported by Nayak \textit{et al.}~\cite{nayak2005hollow}.

\textit{Core preparation.} N-isopropylacrylamide (NIPAM, 0.5 g), 5 mol \% methacrylic acid (MAA) and 10 mol \% N,N’-(1,2-dihydroxyethylene)bisacrylamide (DHEA) were dissolved in 50 mL MQ water at room temperature. The reaction mixture was then immersed into an oil bath at 80 °C and purged with nitrogen for 1 h. The reaction was started by adding 6.5 mg of potassium persulfate (KPS) previously dissolved in 1 mL MQ water and purged with nitrogen. The polymerization was carried out for 6 h in a sealed flask. Afterwards, the colloidal suspension was cleaned by dialysis for a week, and 8 centrifugation cycles and resuspension in pure water. The use of a lower amount of initiator and the absence of surfactants in the reaction mixture allowed to produce microgels of bigger size in comparison to Ref.~\cite{nayak2005hollow}.

\textit{Shell addition.} In a reaction flask, 29 mg of the core microgels were dispersed in 10 mL MQ water at 80 °C and purged with nitrogen for 1 h. In parallel, NIPAM (0.1 g), 5 mol \% MAA and 5 mol \% N,N'-Methylenebis(acrylamide) (BIS) were dissolved in 10 mL MQ water at room temperature and purged with nitrogen for 1 h. In a third vial, 2 mg of KPS were dissolved in 1 mL MQ water and purged with nitrogen. Still keeping the flask sealed, the temperature of the solution containing the core particles was raised to 80°C prior to add the initiator. Immediately afterwards, we started feeding of the monomer solution (at 166 µL/min) into the reaction flask. When the feeding was terminated, the reaction was quenched by opening the flask in air and placing it in an ice bath. The obtained colloidal suspension was cleaned by dialysis for a week, and 8 centrifugation cycles and resuspension in pure water. A polymerization reaction by continuous monomer addition was chosen over the more common batch reaction in order to ensure a more controlled and homogeneous shell growth.\cite{Acciaro2011,Still2013}

\textit{Core degradation.} The controlled degradation of the particles’ cores was achieved by exposing core-shell particles to known amounts of sodium periodate (NaIO$_4$) over different times. In particular, the required amount of mols of sodium periodate with respect to DHEA molecules in solution was added to 500 µL of microgel suspension, and the total volume was raised to 1.5 mL. The amount of DHEA was estimated knowing the microgel concentration, the cores size, and assuming that the entire amount of DHEA added during the cores synthesis was incorporated in the particles. Three different exposure times were used: 20h, 3 days and 8 days. When the required time elapsed, the particle suspension was cleaned by centrifugation and supernatant exchange once a day for 10 days in order to ensure complete removal of loose polymer chains from the interior of the particles.

\subsection{Experimental methods}

\textit{DLS and SLS.} Dynamic light scattering (DLS) experiments were performed using a Zetasizer (Malvern, UK). The samples were let to equilibrate for 15 min at the required temperature (22 or 40°C) prior to performing six consecutive measurements. To record volume phase transition curves, the temperature was scanned from 20 to 50°C with 2°C steps. At each temperature the sample was let to equilibrate for 10 min before performing four consecutive measurements. For static light scattering (SLS), a CGS-3 Compact Goniometer (ALV, Germany) system was used, equipped with a Nd-YAG laser, $\lambda = 532$ nm, output power 50 mW before optical isolator, measuring angles from 30° to 150° with 2° steps.

\textit{Form factor fitting procedure.} Static scattering form factor analysis was performed using the FitIt! tool developed by Otto Virtanen for MATLAB \cite{Virtanen2016}. The intensity distributions were fitted by assuming the following density profile:

$$\rho(r)=\begin{cases}
               \rho_0 e^{\lambda\left(r-R\right)},~0\leq r\leq R\\
               0,~r > R
            \end{cases}$$

This minimal model gives an excellent description of the experimental form factors of the hollow microgels (see Figure S1) with only two parameters: the size $R$ and the rate of decay of the density from the outer surface toward the centre of the microgel $\lambda$. The size of the microgels is assumed to be normally distributed. 

To calculate the mass removed during the reaction with NaIO$_4$: $\Delta M=M_0-M$, we calculated the final mass as $M= m \int_0^R 4 \pi r^2 \rho(r) dr$ and the initial mass as $M_0=4/3\pi R^3 m \rho_0$, where $m$ is the mass of a monomer unit. We then calculated the relative mass loss $\frac{\Delta M}{M_0}$ as:
\begin{equation}
    \frac{\Delta M}{M_0}= 1-3\frac{\lambda R \left(\lambda R-2\right) +2\left(1-e^{-\lambda R}\right)}{\lambda^3R^3}
\end{equation}
We then estimated the relative mass loss of the core $X$ as:
$\frac{M_{c0}-M_c}{M_{c0}}\simeq \frac{\Delta M}{M_0}\left(\frac{M_0}{M_{c0}}\right)=\frac{\Delta M}{M_0}\left(\frac{R^3}{R_{c}^3}\right)$. Where $M_{c}$ and $M_{c0}$ are the final and the initial mass of the core, and $R_{c}$ is the radius of the core.

The 95\% confidence intervals of $X$ were estimated using the Monte Carlo method. In particular, the lower and upper bounds were constructed by calculating the 5$^{th}$ and the 95$^{th}$ percentile of the distribution of $X$ obtained by propagating the uncertainty in $R$, $R_c$, $\lambda$ after 10000 trials. The values of $X$ plotted in Figure \ref{fig:Figure1} as symbols are the medians of such distributions.

\textit{Deposition of isolated microgels from a water/hexane interface.} Microgels were deposited from a water/hexane interface onto silicon wafers for atomic force microscopy (AFM) imaging following an already reported procedure.\cite{Isa2010} Silicon wafers were cut into pieces and cleaned by 15 min ultrasonication in toluene (Fluka Analytical, 99.7\%), isopropanol (Fisher Chemical, 99.97\%), acetone, ethanol and MQ water. A piece of silicon wafer was placed inside a Teflon beaker on the arm of a linear motion driver and immersed in water. Successively, a liquid interface was created between MQ water and n-hexane (Sigma-Aldrich, HPLC grade 95\%). Around 100 µL of the microgels suspension was injected at the interface after appropriate dilution in a 4:1 MQ-water:IPA solution. After 10 min equilibration time, extraction of the substrate was conducted at a speed of 25 µm·s$^{-1}$ to collect the microgels by sweeping through the liquid interface.

\textit{Langmuir trough deposition.} Microgels self-assembled at the water/hexane interface at controlled surface pressure ($\Pi$) values were deposited onto silicon wafers for visualization using a custom-made setup already reported in literature~\cite{rey2016isostructural}. We used a KSV5000 Langmuir trough equipped with a dipper arm immersed in water for holding a silicon substrate forming an angle of approximately 30° with the water surface. The silicon substrate was further cleaned in a UV-Ozone cleaner (UV/Ozone Procleaner Plus, Bioforce Nanosciences) for 15 min to ensure a hydrophilic surface prior to microgel deposition. After forming a water/hexane interface, the substrate was lifted so to pierce the liquid interface. Microgels were then injected on the liquid interface while the surface pressure was simultaneously measured with a platinum Wilhelmy plate. When the required initial surface pressure was reached, the injection was stopped and the interface was left to equilibrate for 15 min. Successively, the dipper was activated to extract the substrate at a constant speed of 0.3 mm·min$^{-1}$ and, after 2 min, the barriers started moving at a compression speed of 2.3 mm·min$^{-1}$. When the compression finished, the barriers were immediately opened while the substrate was still moving up in order to achieve a discontinuity in microgel concentration deposited on the silicon wafer. 

The conformation of microgels at the interface and their 2D assembly as a function of the surface pressure was then inferred by analysing the substrates using atomic force microscopy (AFM). Images from the initial position of the three-phase contact line to the end of the substrate were recorded at a fixed distance of 500 µm. The discontinuity in microgels deposition ensures a correct assignment of the surface pressure value measured at the liquid interface during compression to the corresponding position on the silicon substrate. More specifically, the highest value of surface pressure measured during the experiment was assigned to the position on the substrate corresponding to the highest density of microgels. Consequently, knowing the dipper speed and the distance between AFM images of the substrate, the surface pressure curve was scanned backwards assigning to each AFM image its corresponding value of $\Pi$.

\textit{AFM imaging and analysis.} Microgels deposited on silicon wafers were characterized by AFM (Bruker Icon Dimension), in tapping mode, using cantilevers with 300 kHz resonance frequency and 26 mN·m$^{-1}$ spring constant. Height and phase images were recorded at the same time. Images were first processed with Gwyddion and successively analysed with custom MATLAB codes. The following procedure was used to obtain an average height profile for each microgel: for each microgel, horizontal and vertical profiles passing through its center were measured from AFM height images. Successively, an average over around 10 microgels was obtained by aligning each profile by its center value. For the height profiles of a single microgel inside 2D assemblies (Fig.~\ref{fig:collectivemgelinterf}), the same procedure was used; the profiles were then cut on the x-axis to exclude neighbouring microgels.

The average inter-particle distance $d_{cc}$ at different $\Pi$ was estimated by extracting the positions of the microgels from AFM images taken at different locations on the substrates. For a given set of particles' coordinates (x,y), $d_{cc}$ was calculated as the average distance between neighbouring particles. The neighbours' list was constructed based on the Voronoi tessellation using the Freud open-source Python libraries~\cite{eric_s_harper_2016_166564}. Such neighbours' list was also used to calculate the average hexatic order parameter parameter $\psi_6$ :
\begin{equation}
  \psi_6=\left\langle \frac{1}{N_j}\sum_{k=1}^{N_j} e^{i6\theta_{jk}}\right\rangle
\end{equation}
 $N_j$ is the number of neighbours of the j-th particle in the AFM image, $\theta_{jk}$ is the angle between the unit vector (1,0) and the vector $\textbf{r}=\textbf{r}_k-\textbf{r}_j$ connecting particle $j$ and its k-th neighbour. 

The compression isotherms ($\Pi=f(d_{cc})$) were fitted by assuming that i) the interparticle interactions between microgels adsorbed at the oil-water interface can be described by the generalized Hertzian potential~\cite{RN3707,RN3691}, and that ii) the microgels assemble into a hexagonal lattice~\cite{Grillo2020}:
\begin{equation}\label{eq:GHP}
    \Pi|_{T=0}=-\frac{\partial E_{hex}}{\partial A_{hex}}=\frac{\epsilon\sqrt{3}}{a s r}\left(1-\frac{r}{a}\right)^{a-1}\Theta\left(1-\frac{r}{s}\right)
\end{equation}
where $s$ is the diameter of the isolated microgels, $\epsilon$ is the energy scale of the interactions, $\Theta$ is the Heaviside step function and $a$ the power-law exponent defining the softness of the potential. The fitting was performed using the curve fitting tool of Matlab. A good description of the experimental data was obtained by constraining the range of $d_{cc}$ to 0.75-1 $s$. This is because for $d_{cc}\geq s$ the monolayers are not hexagonally packed, and for $d_{cc}\lesssim 0.75 s$, $\Pi$ displays a sudden change of slope that cannot be captured by the functional form of the generalized Hertzian potential.

\subsection{Numerical models and methods}

\textit{In silico synthesis.} The coarse-grained microgels used for the simulations are characterized by a fully-bonded, disordered network, as previously described in Refs.~\cite{ninarello2019modeling,gnan2017silico}. This is obtained starting from an ensemble of $N$ two- and four-folded patchy particles, which mimic the connectivity of monomers and crosslinkers employed in the chemical synthesis. The spherical shape of the microgels is obtained by letting the patchy particles  self-assemble in a spherical cavity of radius $Z$. For the numerical synthesis of hollow microgels, we also apply an inner spherical force field, which does not allow patchy particles to enter a region of radius $Z_{in}$, with $Z_{in}<Z$. For standard microgels, we add a designing force on crosslinkers only in order to reproduce their inhomogeneous distribution between core and outer periphery as detailed in Ref.~\cite{ninarello2019modeling}. For the hollow microgels investigated here, there are no major differences in applying or not the additional designing force (Fig. S10). Single microgels are assembled with the \textsc{oxdna} simulation package~\cite{oxDNA_edge}.

The assembly process is terminated once almost all possible bonds are formed ($> 99.9\%$). Subsequently, the topology of the microgel is fixed by means of the Kremer-Grest bead spring model~\cite{kremer1990dynamics}, according to which all particles experience a steric repulsion via the Weeks-Chandler-Anderson (WCA) potential,
\begin{equation}
V_{\rm WCA}(r)=
\begin{cases}
4\epsilon\left[\left(\frac{\sigma}{r}\right)^{12}-\left(\frac{\sigma}{r}\right)^{6}\right] + \epsilon & \text{if $r \le 2^{\frac{1}{6}}\sigma$}\\
0 & \text{otherwise,}
\end{cases}
\end{equation}
where $\epsilon$ sets the energy scale, $\sigma$ is the diameter of the particles, which also defines the unit of length, and $r$ is the distance between two particles. Also, bonded particles interact via the Finitely Extensible Nonlinear Elastic (FENE) potential,
\begin{equation}
V_{\rm FENE}(r)=-\epsilon k_FR_0^2\ln\left[1-\left(\frac{r}{R_0\sigma}\right)^2\right]     \text{ if $r < R_0\sigma$,}
\end{equation}
with $k_F=15$ which determines the stiffness of the bond and $R_0=1.5$ is the maximum bond distance. In this work, we study microgels with $Z=100\sigma$ and $75\sigma$ with a fraction of crosslinker $c=5\%$.
The model for standard microgels employed in this work is the one described in Ref.~\cite{ninarello2019modeling}, with an average internal density of monomers $\rho\sim 0.08 \sigma^{-3}$. For the hollow particles, we employ $Z_{in}=0.75Z$ and $\rho=0.035 \sigma^{-3}$; the consequences of varying these parameters are discussed in the Supporting Information file, Figs. S7-S10. In all cases, we assume that the low amount of charges present after the chemical synthesis does not affect the structural features and the swelling behavior of the \textit{in silico} microgels.

The protocol just described does not allow to reproduce all the stages of the chemical synthesis procedure. Nonetheless, it allows generating standard and hollow microgels independently, ensuring that the structural features of the final particles resemble experimental results. For this reason, in order to assess the core degradation process, we also make use of a standard microgel with $Z=90\sigma$, $\rho=0.08\sigma^{-3}$ and $c=10\%$ (corresponding to the DHEA-crosslinked microgel in the chemical synthesis), which is then inserted in the central cavity of a hollow microgel, mimicking a core-shell particle. By progressively removing the internal monomers and thus reducing the internal density, we reproduce the experimentally accessible core degradation stages (Fig. S11).
Finally, in order to provide a qualitative comparison between the form factors extracted during this process and the experimental ones, we need to further consider the experimentally observed size increase as a function of $X$, and accordingly rescale each numerical form factor to obtain a comparison in nm$^{-1}$.

\textit{Bulk behavior.} Microgels in bulk are investigated by means of molecular dynamics simulations with implicit solvent. Thermoresponsivity is mimicked by adding a solvophobic potential 
\begin{equation}\label{eq:valpha}
V_{\alpha}(r)  =  
\begin{cases}
-\epsilon\alpha & \text{if } r \le 2^{1/6}\sigma  \\
\frac{1}{2}\alpha\epsilon\left\{\cos\left[\delta{\left(\frac{r}{\sigma}\right)}^2+\beta\right]-1\right\} & \text{if } 2^{1/6}\sigma < r \le R_0\sigma  \\
0 & \text{if } r > R_0\sigma
\end{cases}
\end{equation}
with $ \delta = \pi \left(\frac{9}{4}-2^{1/3}\right)^{-1} $ and $\beta = 2\pi - \frac{9}{4}\delta$~\cite{soddemann2001generic}. $V_\alpha$ effectively introduces an attraction among all monomers of the network, modulated by the parameter $\alpha$: in case $\alpha=0$, the standard Kremer-Grest model is recovered whereas higher values of $\alpha$ mimic an increasing temperature of the dispersion, leading to a microgel in the collapsed state at $\alpha \approx 1.1$. Simulations in bulk are performed in the $NVT$ ensemble, fixing the reduced temperature $T^*=k_BT/\epsilon=1$ with the Nosé-Hoover thermostat.

\textit{Interfacial behavior.} Investigations at the liquid-liquid interface are carried out in the presence of explicit solvent particles in order to reproduce the effect of the surface tension between the two solvents. Solvent particles are modeled as soft beads within the dissipative particle dynamics (DPD) framework~\cite{groot1997dissipative,camerin2018modelling}. The total interaction force among beads is $\vec{F}_{ij} = \vec{F}^C_{ij} + \vec{F}^D_{ij} + \vec{F}^R_{ij}$, where:
\begin{eqnarray}
	\vec{F}^C_{ij}  &=&  a_{ij} w(r_{ij}) \hat{r}_{ij} \\
	\vec{F}^D_{ij}  &=&  -\gamma w^2(r_{ij}) (\vec{v}_{ij}\cdot\vec{r}_{ij}) \hat{r}_{ij} \\
	\vec{F}^R_{ij}  &=&  2\gamma\frac{k_B T}{m} w(r_{ij}) \frac{\theta}{\sqrt{\Delta t}} \hat{r}_{ij}
\end{eqnarray}
where $\vec{F}^C_{ij}$ is a conservative repulsive force, with $w(r_{ij}) = 1-r_{ij}/r_c$ for $r_{ij}<r_c$ and $0$ elsewhere, $\vec{F}^D_{ij}$ and $\vec{F}^R_{ij}$ are a dissipative and a random contribution of the DPD, respectively; $a_{ij}$ quantifies the repulsion between two particles, $\gamma=2.0$ is a friction coefficient, $\theta$ is a Gaussian random variable with zero average and unit variance, and $\Delta t=0.002$ is the integration time-step. According to previous works~\cite{camerin2019microgels,camerin2020microgels}, in order to reproduce a water/hexane (w/h) interface, we choose $a_{\rm ww}=a_{\rm hh}=8.8$, $a_{\rm hw}=31.1$. Instead, for the monomer-solvent interactions we choose $a_{\rm mw}=4.5$ and $a_{\rm mh}=5.0$.  The cut-off radius is always set to be $r_c=1.9 \sigma$ and the reduced solvent density $\rho_{\rm DPD}=4.5$. In this case, the reduced temperature $T^*$ is fixed to $1$ via the DPD thermostat. 
At the interface, we limit the study to microgels with $Z=75\sigma$, because of the exceptional computational cost to carry out the simulations in the presence of explicit solvent. Under these conditions, more than $5 \times 10^6$ solvent particles are inserted in the simulation box. Simulations are performed with the \textsc{lammps} simulation package~\cite{plimpton1995fast}.

For microgels adsorbed at the interface, we perform compression tests by imposing an external force of cylindrical symmetry, with the main axis perpendicular to the plane of the interface, along the $z$ axis. In this way, microgel monomers experience an harmonic force $F(r)=-k(r-R)^2$ with $F(r)=0$ if $r>R$, where $r$ is the distance from the monomer to the center axis of the cylinder, $R$ is the equilibrium radius of the cylinder and $k=10$ is the intensity of the force. Solvent particles are not subjected to $F(r)$.

\textit{Measured quantities.} In order to compare simulations and experiments, we calculate the numerical form factors in bulk as 
\begin{equation}
P(q)=\frac{1}{N}\sum_{i,j=1}^N \langle \exp{(-i\vec{q} \cdot \vec{r}_{ij})} \rangle,
\end{equation} 
where $r_{ij}$ is the distance between monomers $i$ and $j$, while the angular brackets indicate an average over different configurations and over different orientations of the wavevector $\vec{q}$. In real space, we calculate the radial density profiles of the microgel as
\begin{equation}\label{eq:profile}
\rho(r)= \left\langle \frac{1}{N}\sum_{i_{=1}}^{N} \delta (|\vec{r}_{i}-\vec{r}_{CM}|-r) \right\rangle.
\end{equation} 
with $\vec{r}_{CM}$ the distance from the microgel center of mass. At the interface, similarly to Ref.~\cite{camerin2019microgels}, we also determine $\rho(z)$, which is the density profile obtained by dividing the simulation box along the $z$ axis (being $z$ perpendicular to the interfacial plane) into three dimensional bins that are parallel to the interface, and $\rho(\zeta)$, with $\zeta=x,y$, for which bins are taken orthogonally to the interfacial plane. The latter is calculated at a distance $\zeta$ from the microgel center of mass and averaged over $x$ and $y$. 

\normalsize
\section{Acknowledgments}
We thank Miguel Angel Fernandez-Rodriguez for  discussions. J.V. and L.I. acknowledge Dr. Kirill Feldman and Prof. Jan Vermant for instrumentation access and discussion, and Dr. Shivaprakash Narve Ramakrishna for technical support.
J.V. acknowledges funding from the European Union’s Horizon 2020 research and innovation programme under the Marie Skłodowska Curie grant agreement 888076.
F.C., L.R. and E.Z. acknowledge financial support from the European Research Council (ERC Consolidator Grant 681597, MIMIC) and gratefully acknowledge the computing time granted by EUSMI on the supercomputer JURECA at the Jülich Supercomputing Centre (JSC).

\section*{Author Contributions}
Author contributions are defined based on the CRediT (Contributor Roles Taxonomy) and listed alphabetically. Conceptualization: L.I. and E.Z.; Formal analysis: F.C., F.G. and J.V.; Funding acquisition: L.I., J.V. and E.Z.; Investigation: F.C., F.G., L.I., J.V. and E.Z.; Methodology: F.C., F.G., L.R., J.V. and E.Z.; Project administration: L.I. and E.Z.; Software: F.C., F.G. and L.R.; Supervision: L.I and E.Z.; Validation: F.C. and J.V.; Visualization: F.C., F.G., L.I., J.V. and E.Z.; Writing – original draft: F.C., F.G., L.I., J.V. and E.Z.; Writing – review and editing: F.C., F.G., L.I., L.R. J.V. and E.Z..


\clearpage
\newpage
\begin{center}
\Large
\textbf{The effect of internal architecture \\on the assembly of soft particles at fluid interfaces\\ \bigskip \large Supporting Information}

\normalsize
\bigskip
Jacopo Vialetto\textsuperscript{ *,1}, Fabrizio Camerin\textsuperscript{ *, 2, 3}, Fabio Grillo\textsuperscript{ 1},\\ Lorenzo Rovigatti\textsuperscript{ 4, 2}, Emanuela Zaccarelli\textsuperscript{ *, 2, 4}, Lucio Isa\textsuperscript{ *, 1}\\
\medskip
\small
\textit{%
\textsuperscript{1}Laboratory for Soft Materials and Interfaces, Department of Materials, ETH Z{\"u}rich, Vladimir-Prelog-Weg 5, 8093 Z{\"u}rich, Switzerland\\
\textsuperscript{2}CNR Institute of Complex Systems, Uos Sapienza, Piazzale Aldo Moro 2, 00185 Roma, Italy\\
\textsuperscript{3}Department of Basic and Applied Sciences for Engineering,\\ Sapienza University of Rome, via Antonio Scarpa 14, 00161 Roma, Italy\\
\textsuperscript{4}Department of Physics, Sapienza University of Rome, Piazzale Aldo Moro 2, 00185 Roma, Italy
}

Email: jacopo.vialetto@mat.ethz.ch; fabrizio.camerin@gmail.com; emanuela.zaccarelli@cnr.it; lucio.isa@mat.ethz.ch

\end{center}

\renewcommand{\thefigure}{S\arabic{figure}}\setcounter{figure}{0}

\newpage

\section{Supplementary Experimental Data}

\begin{figure}[!htb]
\centering
\includegraphics[scale=0.95]{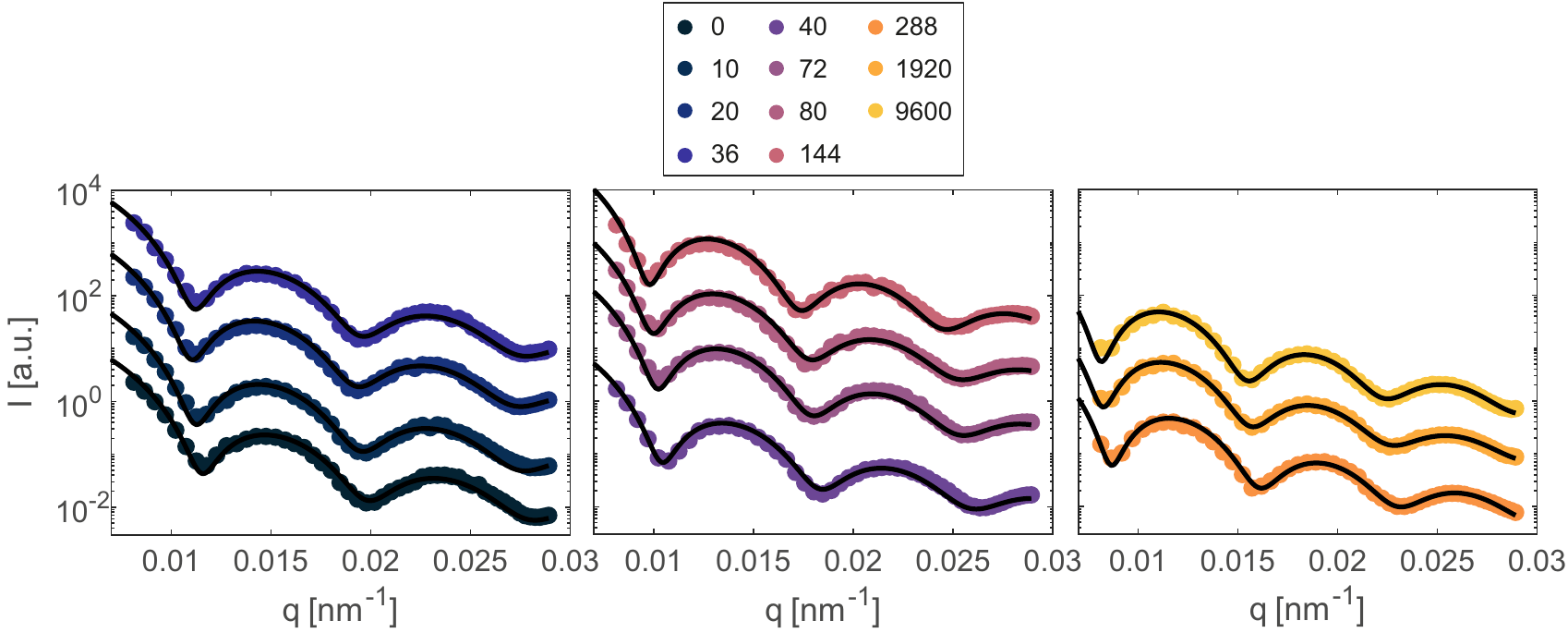}
\caption{\small \textbf{Form factors of core-shell microgels as function of core removal.} Experimental form factors obtained from SLS experiments at 25°C as function of the core degradation parameters $n \cdot t$ reported above the graphs, arbitrarily rescaled on the y-axis for clarity. Black lines are fits (see Methods).}
\label{fig:allformfactors}
\end{figure}

\begin{figure}[!htb]
\centering
\includegraphics[scale=0.9]{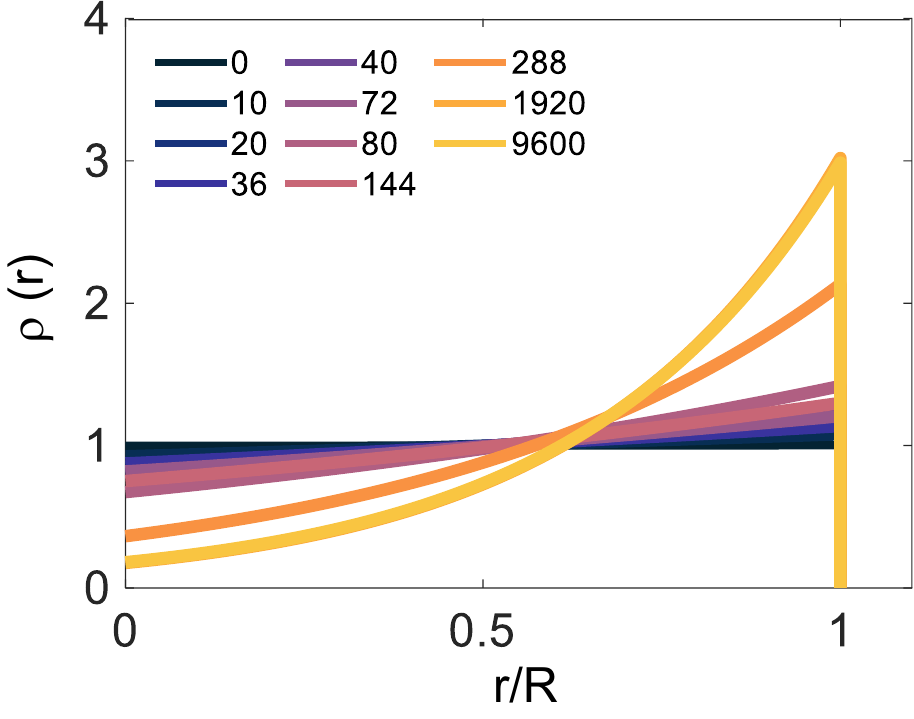}
\caption{\small \textbf{Density profiles of core-shell microgels as function of core removal.} Microgel radial density profiles ($\rho(r)$) plotted as a function of a normalized radial coordinate $r/R$, where $R$ is the particle radius, as extracted from the fitting procedure of the experimental form factors, as explained in the Methods section.}
\label{fig:densproffromfit}
\end{figure}

\begin{figure}[!htb]
\centering
\includegraphics[scale=0.7]{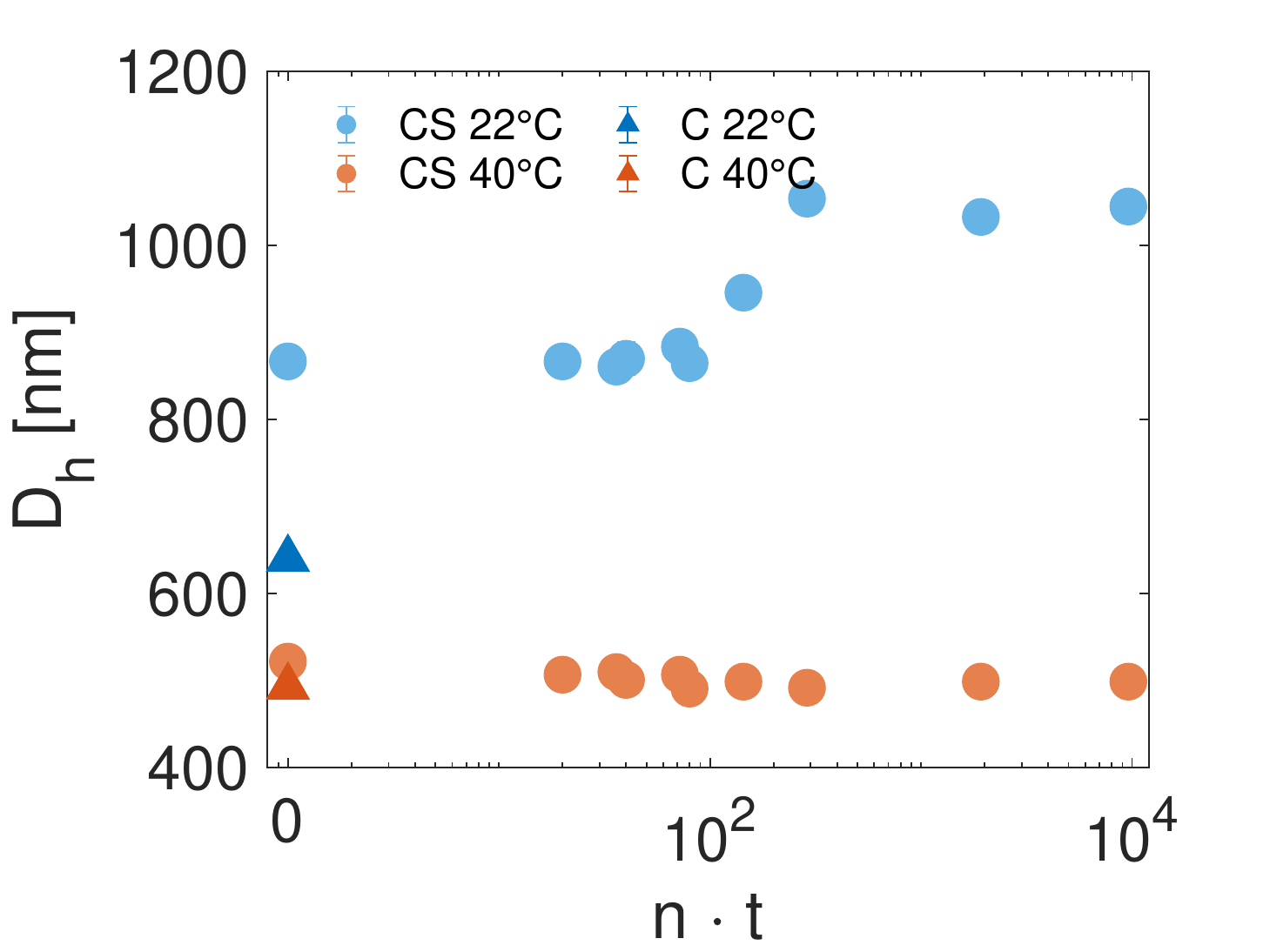}
\caption{\small \textbf{Microgels' characterization by DLS.} Experimental hydrodynamic diameter $D_h$ for the core (triangles) and core-shell microgels as a function of the core degradation process (circles), expressed as $n \cdot t$,  at 22°C and 40°C. Error bars indicate the standard deviation of 4 measurements consisting of 13 runs each and are smaller than the symbols size.}
\label{fig:dls}
\end{figure}

\begin{figure}[!htb]
\centering
\includegraphics[scale=0.7]{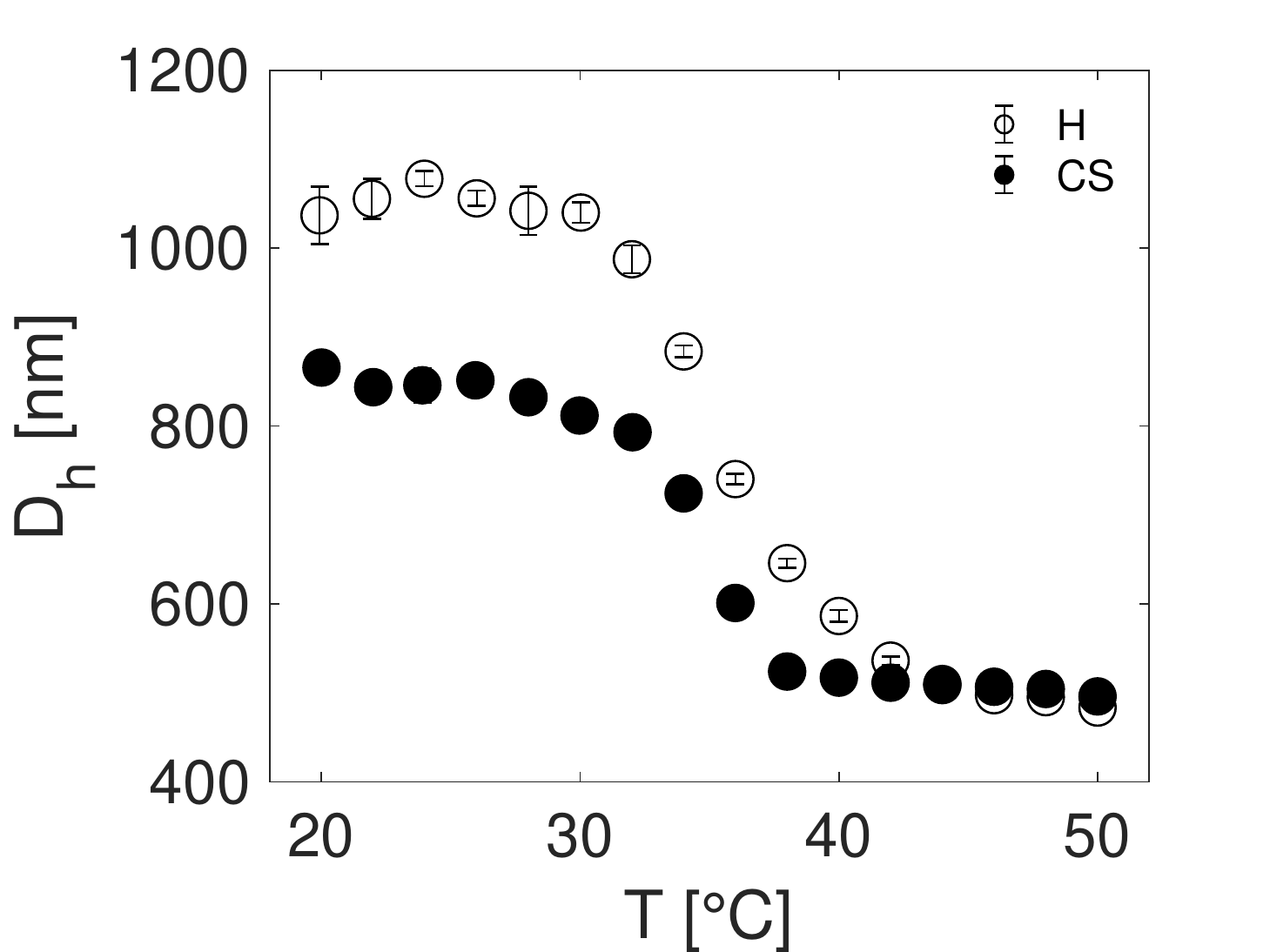}
\caption{\small \textbf{Hydrodynamic diameter vs temperature.} Hydrodynamic diameter $D_h$ for core-shell (filled symbols, corresponding to $X=0\%$) and hollow (hollow symbols, $X\simeq100\%$) as function of the solution temperature. Error bars indicate the standard deviation of 4 measurements consisting of 13 runs each and are smaller than the symbols size.}
\label{fig:DLS_Tscan}
\end{figure}


\begin{figure}[!htb]
\centering
\includegraphics[scale=1]{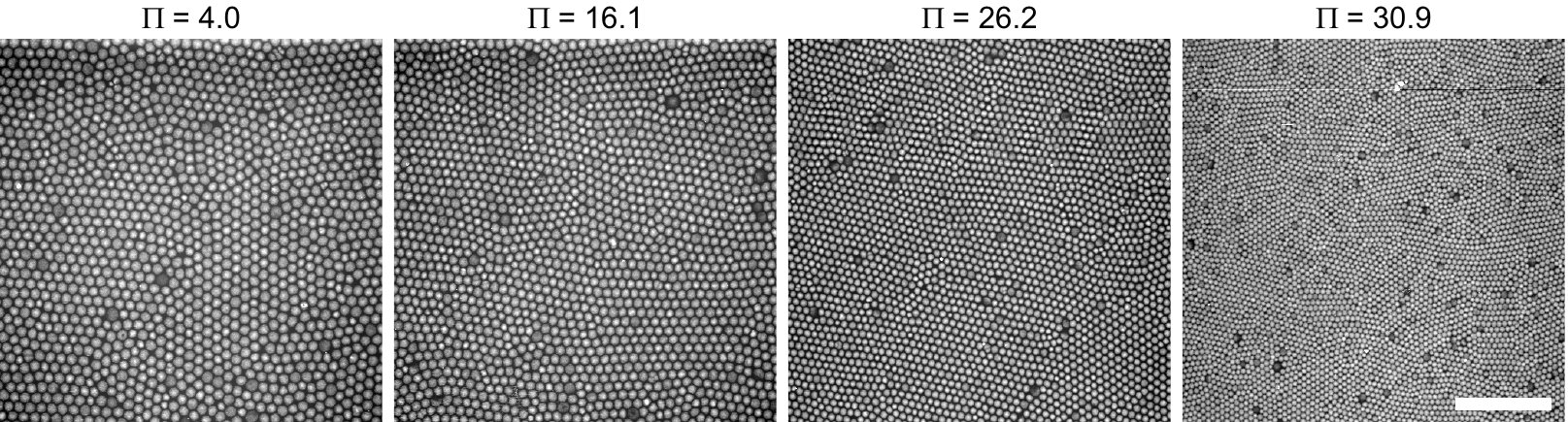}
\caption{\small \textbf{2D assembly of hollow microgels as function of the surface pressure.}  AFM images of the monolayer microstructure after assembly at the water/hexane interface and deposition on the solid substrate, at increasing values of the surface pressure $\Pi$. Scale bar: 20$\mu$m.}
\label{fig:2D_assembly_lowmag}
\end{figure}


\begin{figure}[!htb]
\centering
\includegraphics[scale=0.7]{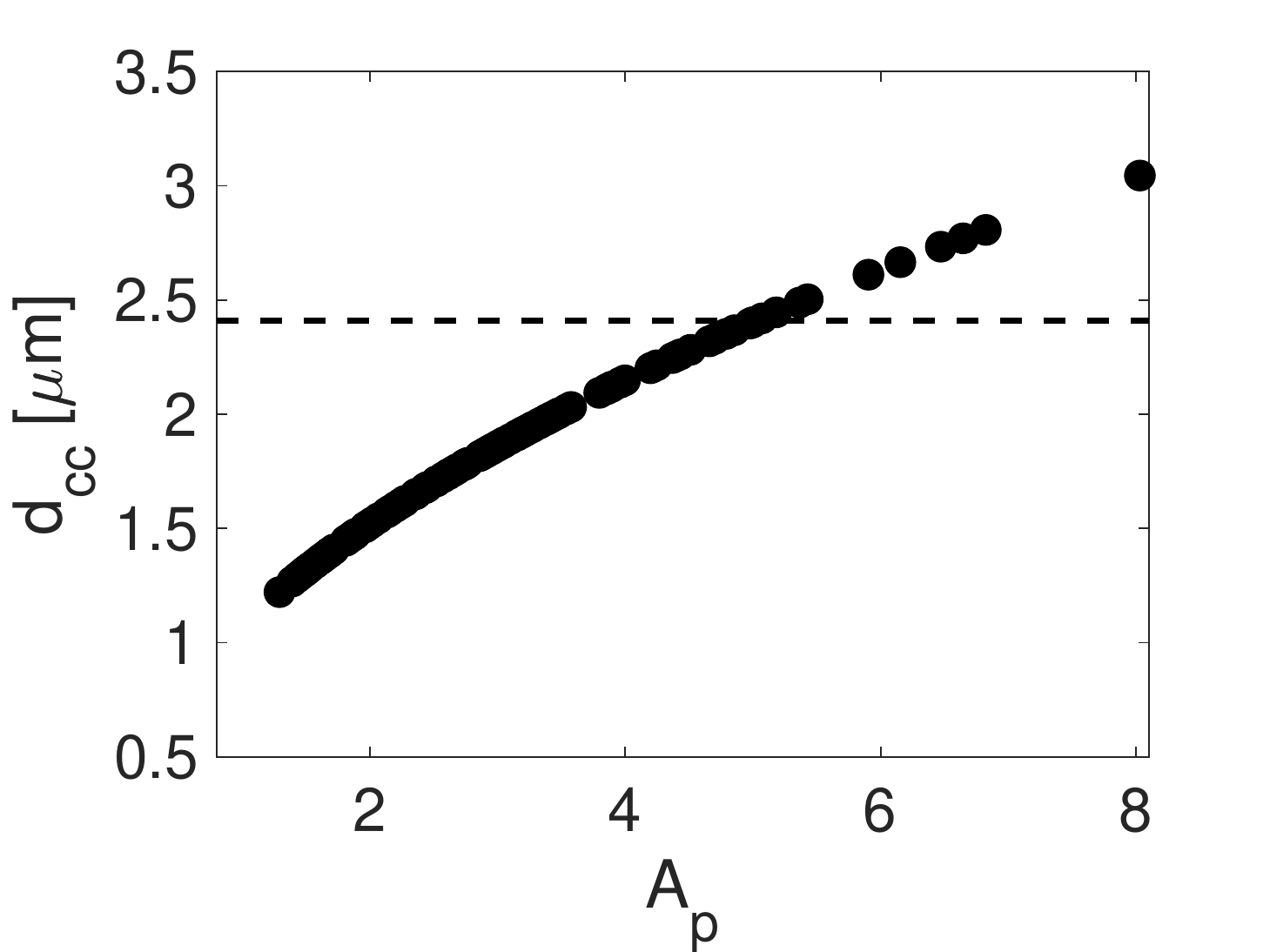}
\caption{\small \textbf{Average center-to-center distance vs area per particle.} Plot of the center-to-center distance $d_{cc}$ as function of the area per particle $A_p$ from images as the ones in \ref{fig:2D_assembly_lowmag}. The surface pressure of the interface increases from left to right. The dashed line indicates the average diameter of a hollow microgel at the interface prior to compression $D^0_{i}$.}
\label{fig:dcc_Ap}
\end{figure}

\clearpage

\section{Supplementary Numerical Methods and Data}

\subsection{Comparison between the numerical form factors}

In this section, we present some of the comparisons we carried out in order to determine the hollow microgel model whose form factor best reproduces the experimental data. As also reported in the main text, the comparison is performed by matching the maximum of the first peak for experimental and numerical form factors and, unless specified, we refer to microgels with $Z=100\sigma$.

\begin{figure}[b!]
\centering
\includegraphics[scale=0.4]{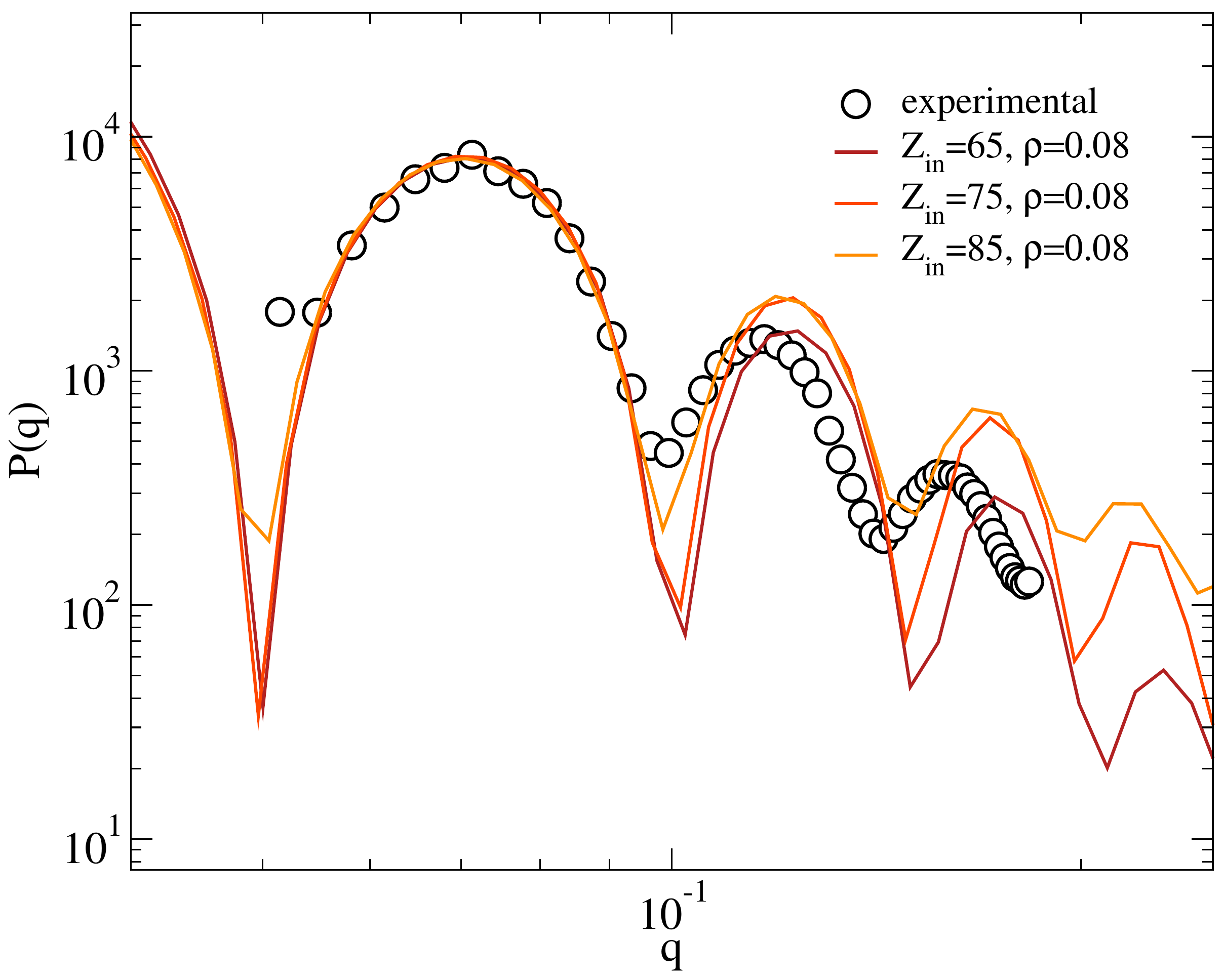}
\caption{\small \textbf{Form factor comparison for the hollow microgels.} Form factors at a fixed number density $\rho=0.08$ and outer radius $Z=100\sigma$, varying the inner size of the cavity $Z_{in}$. Form factors are arbitrarily rescaled in the x and y-axes and compared to experimental data (symbols).}
\label{fig:cfrff_zin_rho008}
\end{figure}

\begin{figure}[t!]
\centering
\includegraphics[scale=0.4]{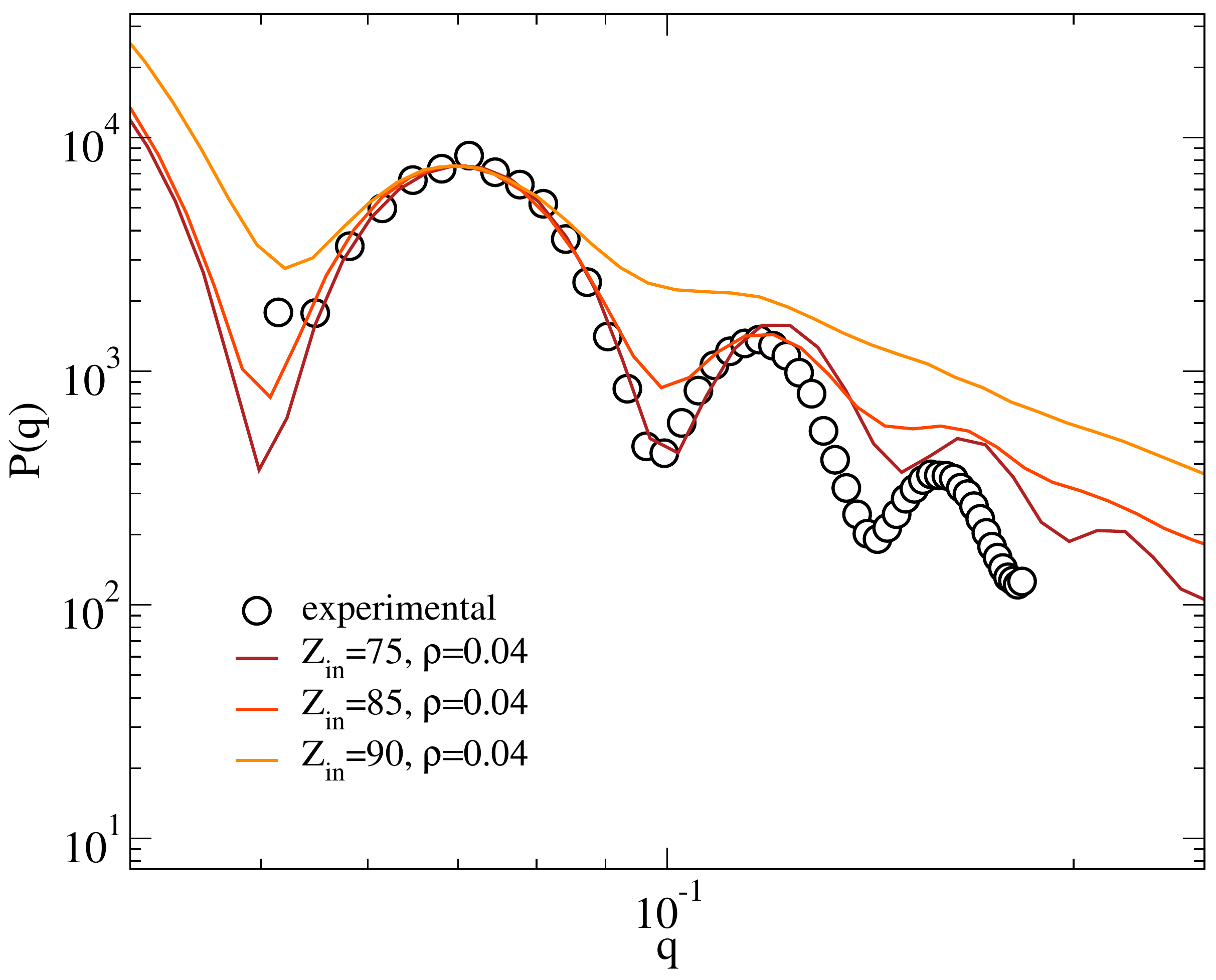}
\caption{\small \textbf{Form factor comparison for the hollow microgels.} Form factors at a fixed number density $\rho=0.04$ and outer radius $Z=100\sigma$, varying the inner size of the cavity $Z_{in}$. Form factors are arbitrarily rescaled in the x and y-axes and compared to experimental data (symbols). }
\label{fig:cfrff_zin_rho004}
\end{figure}

\begin{figure}[h!]
\centering
\includegraphics[scale=0.4]{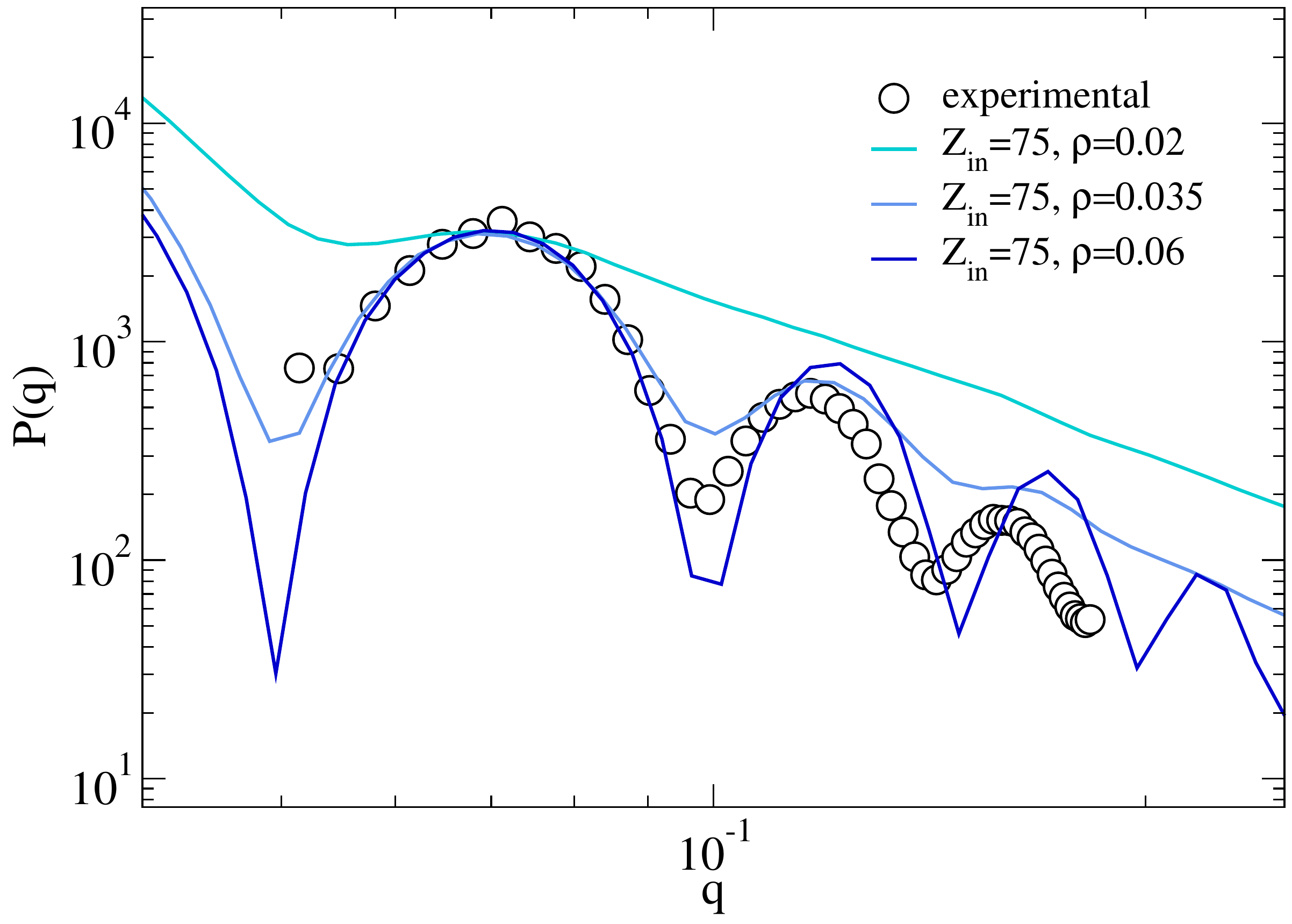}
\caption{\small \textbf{Form factor comparison for the hollow microgels.} Form factors at a fixed inner radius $Z_{in}=75\sigma$ and outer radius $Z=100\sigma$, varying the internal density $\rho$. Form factors are arbitrarily rescaled in the x and y-axes and compared to experimental data (symbols).}
\label{fig:cfrff_zin75_rho}
\end{figure}

First of all, we consider hollow microgels with varying size of the internal hole with the same average monomer density $\rho=0.08$ of a standard core-corona microgel, which was found to favorably compare to experimental form factors~\cite{ninarello2019modeling}. As shown in Fig.~\ref{fig:cfrff_zin_rho008}, despite the fact that the first peak describes the experimental data well, the second and the third peak of the numerical form factors are all shifted to higher wavenumbers. By reducing the internal monomer density to $\rho=0.04$ (Fig.~\ref{fig:cfrff_zin_rho004}), we observe a better agreement, especially for the case with $Z_{in}=75\sigma$. We thus fix the radius $Z_{in}$ and vary $\rho$, as shown in Fig.~\ref{fig:cfrff_zin75_rho}. By reducing $\rho$, we notice that the position of the peaks in terms of the wavenumber is well reproduced by the $\rho=0.035$ case.

\begin{figure}[t!]
\centering
\includegraphics[scale=0.4]{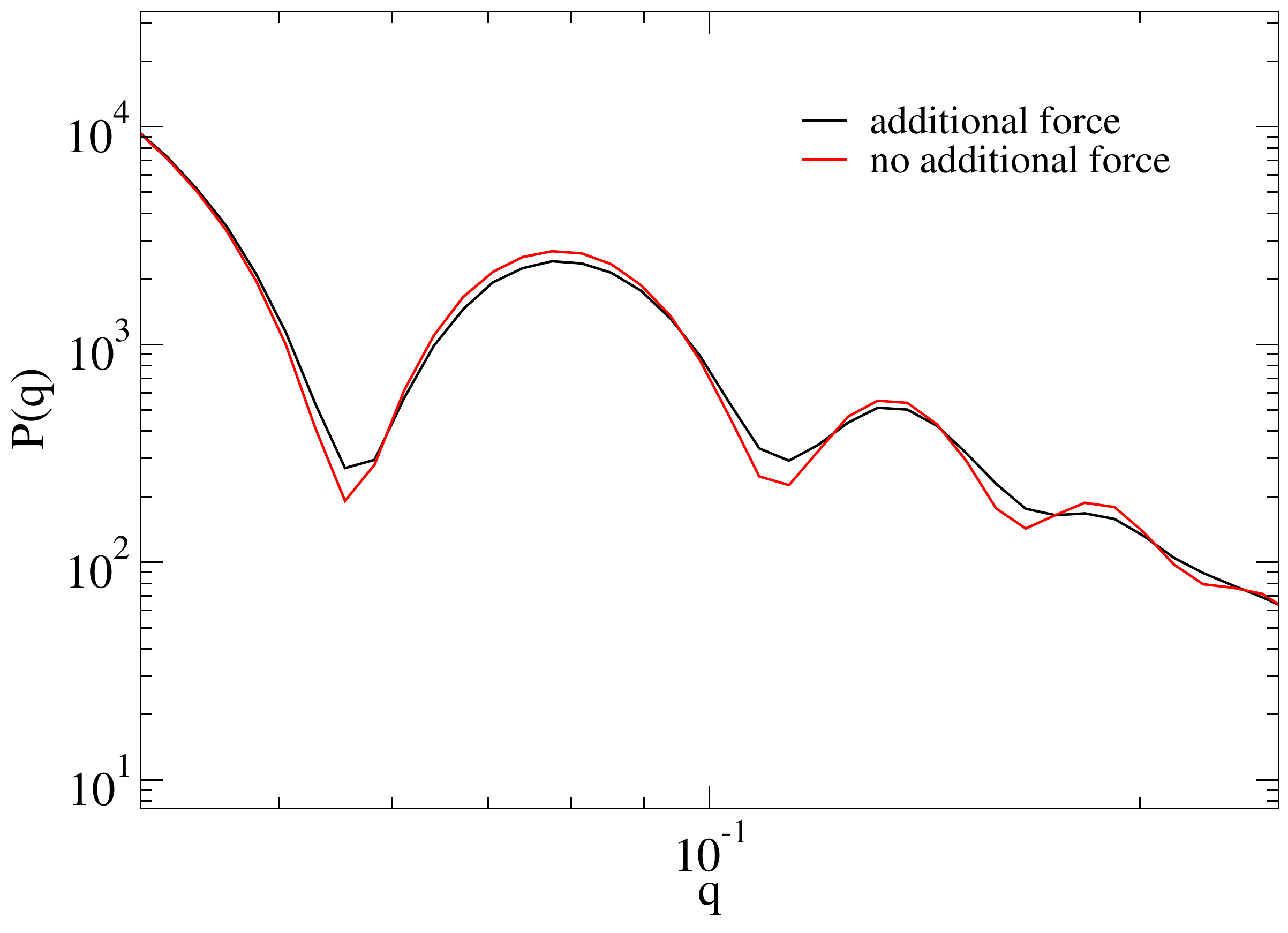}
\caption{\small \textbf{Effect of the designing force on the form factor of hollow microgels.}  The form factors of microgels synthesized with and without a designing force on the crosslinkers are compared. In particular, we consider the case with $Z=100\sigma$, $Z_{in}=75\sigma$ and $\rho=0.035$. Form factors are arbitrarily rescaled in the x and y-axes.}
\label{fig:cfrff_homononhomo}
\end{figure}

Finally, we verify the effect of an additional designing force acting on the crosslinkers. This force was previously introduced in the modeling of standard microgels~\cite{ninarello2019modeling} in order to concentrate a higher amount of crosslinkers in the core of the particle, as it is typically seen in precipitation polymerization synthesis protocols. However, as shown in Fig.~\ref{fig:cfrff_homononhomo}, there is no substantial difference in the form factors for a hollow microgel assembled by employing or not the same designing force used for the standard microgels, and, in particular, the position of the peaks in terms of the wavenumber does not change between the two cases. The minimal differences that remain may be due to small variations in the \textit{in silico} synthesis of the microgels.

Therefore, the hollow microgel model we adopt in the main text is the one with $Z_{in}=75\sigma$ and an average internal density $\rho=0.035$, with no additional designing force applied on the crosslinkers during the assembly process. This choice is also found to describe the deswelling transition of the hollow microgels with increasing temperature appropriately, as shown in Fig.2 of the main text.

\subsection{Numerical simulations of the core degradation process}


\begin{figure}[b!]
\centering
\includegraphics[scale=0.4]{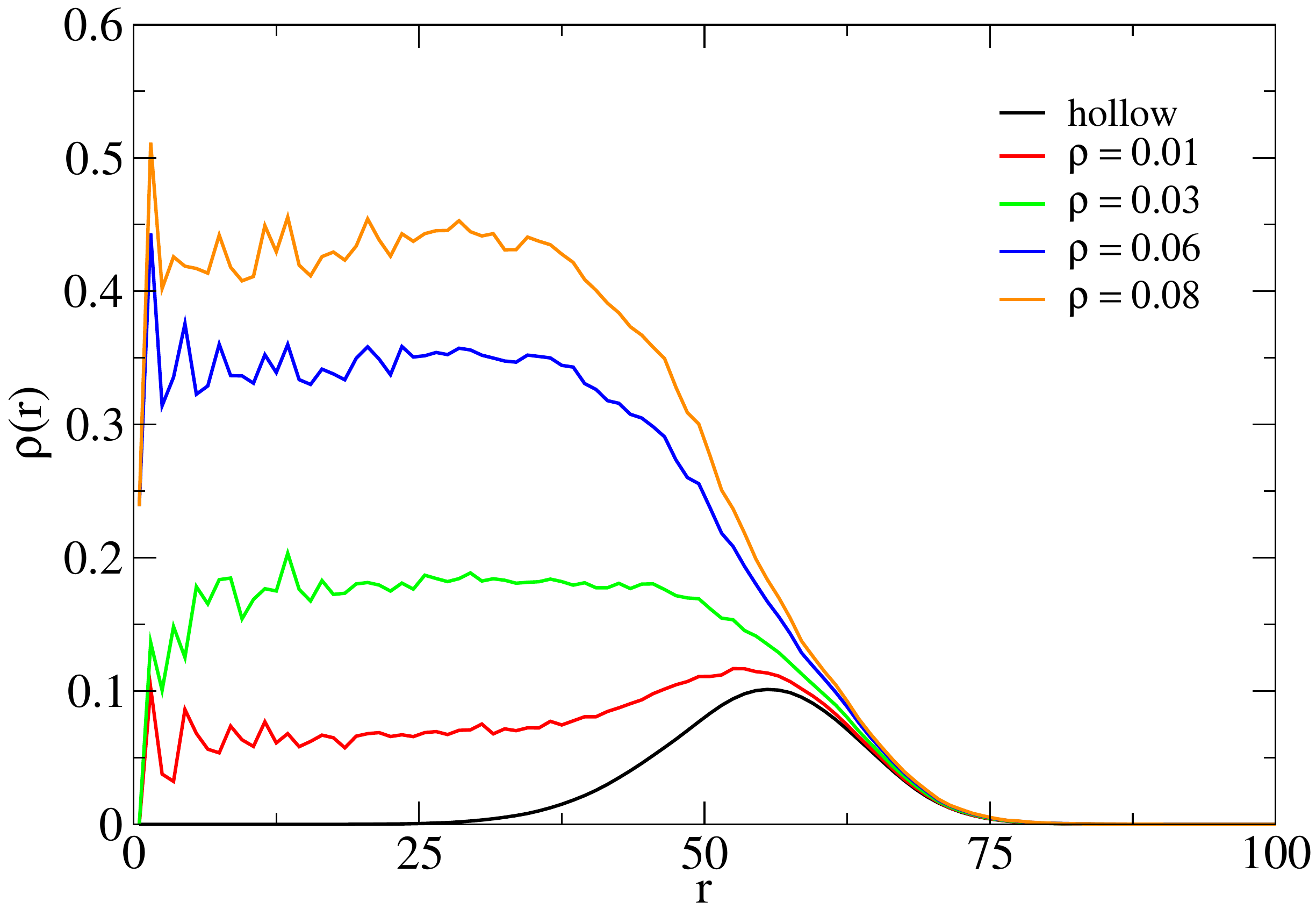}
\caption{\small \textbf{Effect of the core degradation on the core-shell microgel density profiles.} Radial density profiles of the core-shell microgels obtained by inserting a standard microgel inside a hollow one, and subsequently randomly removing monomers from the core microgel in order to obtain a reduced internal density. The initially inserted standard microgel has a density $\rho=0.08$, while the hollow microgel density profile corresponds to $\rho=0$.}
\label{fig:densprofcoredegradation}
\end{figure}

On the numerical side, while it is not possible to faithfully reproduce the chemical stages of the synthesis, we can mimic the core degradation process. In particular, as explained in the main text, we take the hollow microgel whose form factor best reproduces the experimental data and we insert a standard microgel in the central cavity. This corresponds to the experimental state $X = 0\%$, that is the starting core-shell microgel, from which we begin to degrade the core. In simulations, we remove monomers in a random way, progressively reducing the density of the core. Correspondingly,  the numerical form factors, as reported in the main text, show a trend in the position of the peaks that closely compares to the experimental one. Additional information comes from the numerical density profiles in real space: while for experiments these are extracted through a fitting procedure, in simulations they can be directly calculated, see Fig.~\ref{fig:densprofcoredegradation}. It is evident how, starting from a density profile that resembles that of a standard microgel, the progressive removal of monomers leads to the profile of a hollow microgel. Given the correspondence between numerical and experimental form factors, the numerical density profiles constitute a plausible description of the process occurring during the core degradation.

\bibliography{mgel_hollow_bib}

\end{document}